\documentclass[a4paper,11pt]{article}
\usepackage{jheppub} 
\usepackage[utf8]{inputenc}
\usepackage{physics}
\usepackage{booktabs}
\usepackage{bm}
\usepackage[export]{adjustbox}
\newcommand{\pizero}{{\pi^0}}
\newcommand{\piplus}{{\pi^+}}

\usepackage{xcolor}

\begin{document}
\preprint{MITP-26-028,CERN-TH-2026-123}

\author[a,b,c]{Alessandro De Santis,}
\author[d]{Dominik Erb}
\author[a,d,e]{and Harvey B. Meyer}

\affiliation[a]{Helmholtz Institut Mainz, Staudingerweg 18, D-55128 Mainz, Germany}
\affiliation[b]{Johannes Gutenberg-Universit{\"a}t Mainz, 55099 Mainz, Germany}
\affiliation[c]{GSI Helmholtz Centre for Heavy Ion Research, 64291 Darmstadt, Germany}
\affiliation[d]{PRISMA$^+$ Cluster of Excellence \& Institut f{\"u}r Kernphysik, Johannes Gutenberg-Universit{\"a}t Mainz,
D-55099 Mainz, Germany}
\affiliation[e]{Theoretical Physics Department, CERN, 1211 Geneva 23, Switzerland}

\emailAdd{desantia@uni-mainz.de, domerb@uni-mainz.de, h.b.meyer@cern.ch}

\title{Electromagnetic pion mass splitting\\ using a Pauli-Villars-regulated photon propagator}

\abstract{ We present a lattice QCD calculation of the charged–neutral pion mass splitting $M_{\pi^+} - M_{\pi^0}$ at $\order{\alpha_\mathrm{em}}$ using a recently proposed framework based on a Pauli–Villars (PV) regulated photon propagator defined in the continuum and infinite-volume limit, with $\Lambda$ acting as an additional UV cutoff scale. This formulation, when combined with a dedicated treatment of the elastic contribution, avoids power-law finite-volume effects on electromagnetic mass shifts.
We perform the calculation using CLS ensembles, studying finite-volume effects, the continuum limit and the extrapolation to the physical point for several values of the scale $\Lambda$. By means of the Cottingham formula, we further decompose the result into elastic and inelastic contributions at fixed $\Lambda$. Our final result, after removing the cutoff scale $\Lambda$, is $M_{\pi^+} - M_{\pi^0} = 4.56(22)~\text{MeV}$, in good agreement with the experimental measurement. Our complete calculation demonstrates that, for a mass splitting not requiring any counterterms, the formalism can be implemented successfully, whilst also offering useful insights into the computation of electromagnetic corrections to other observables.
}

\date{\today}

\maketitle

\section{Introduction \label{sec::introduction}}

Lattice QCD calculations have reached sub-percent precision for many observables of phenomenological interest. In most cases, computations are performed using a QCD action in the isosymmetric limit, i.e. assuming degenerate light quark masses, $\delta m_\ell = m_d - m_u = 0$, and neglecting electromagnetic effects by setting $\alpha_\mathrm{em} = 0$. However, at the current level of sub-percent precision, these approximations are no longer sufficient for a meaningful comparison with experimental measurements. This is particularly relevant for observables where potential signals of new physics may appear as small effects at or below the percent level.

The inclusion of QED and isospin-breaking effects introduces significant technical challenges, making such calculations computationally demanding and time-consuming. While direct QCD+QED simulations~\cite{RCstar:2022yjz,RC:2025zoa} are actively being pursued, the most widely adopted approach consists in expanding the QCD+QED action around the isosymmetric point in powers of $\alpha_\mathrm{em}$ and $\delta m_\ell$~\cite{deDivitiis:2013xla}, following a long tradition in the subject~\cite{Cottingham:1963zz,Gasser:1974wd,Gasser:1982ap}. Since these parameters are small, they can be treated perturbatively as corrections to correlation functions computed via Monte Carlo simulations using only QCD in the isosymmetric limit. In practice, this corresponds to evaluating diagrams with insertions of electromagnetic vertices connected by a photon propagator.

The implementation of the photon propagator is not unique. As far as its infrared regularization is concerned,  several  schemes have been proposed over time, among which the most commonly used are QED$_\mathrm{L}$~\cite{Hayakawa:2008an}, QED$_{\mathrm{TL}}$ \cite{BMW:2014pzb}, QED$_\infty$~\cite{Asmussen:2016lse}, QED$_C$ \cite{Lucini:2015hfa}, QED$_m$ \cite{Endres:2015gda} and, recently, QED$_r$ \cite{DiCarlo:2025uyj}; we adopt the QED$_\infty$ prescription in the present calculation. In terms of the ultraviolet regularization, the most common choice is to formulate the photon on the lattice, i.e.\ the ultraviolet cutoff for the photon is the same as for the QCD degrees of freedom. As a consequence,
the results obtained with different QCD actions can only be consistently compared after including all necessary counterterms for renormalization and taking the continuum limit, since photon insertions typically induce ultraviolet (UV) divergences. This represents a major drawback, as it complicates comparisons and cross-checks among different lattice collaborations employing different QCD discretizations, as well as comparisons with phenomenological predictions. The issue is especially relevant because lattice QCD calculations often proceed incrementally, determining one class of diagrams at a time. A well-known example is the computation of the leading hadronic contribution to the muon $g-2$, where a solid consensus has been achieved in isoQCD, but not yet for the inclusion of QED and isospin-breaking corrections (see \cite{Aliberti:2025beg} for a review).

A framework aimed at overcoming these limitations has been proposed in~\cite{Biloshytskyi:2022ets}, and this work presents its first application to predicting a hadron mass splitting. The method replaces the photon propagator appearing in QED correction diagrams with a Pauli-Villars (PV) regulated propagator, characterized by a scale $\Lambda$ and defined directly in the continuum and infinite-volume limit. Although this strategy introduces an additional scale $\Lambda\ll a^{-1}$, which must be removed in the limit $\Lambda \to \infty$ after taking the continuum limit, it offers in general several practical advantages:
\begin{itemize}
    \item  At fixed $\Lambda$, no QED-induced counterterm is needed for the continuum limit to exist and for continuum-extrapolated results to be independent of the lattice formulation of QCD; given a common isoQCD scheme, even individual diagrams can rigorously be compared  across collaborations.
    
    \item The cutoff $\Lambda$ makes the propagator a UV-finite `weight function' for QCD correlation functions, and does not induce changes in renormalization and improvement coefficients (the renormalization factor $Z_V$ of the local vector current, for instance).
    
    \item Defining the propagator in infinite volume (with or without a PV mass) avoids the power-law finite-volume effects associated with the photon in a periodic box that arise, for example, in the QED$_L$ framework. This is particularly advantageous for quantities such as the hadronic vacuum polarization and the hadronic light-by-light contribution to the muon $g-2$. However, for charged-hadron self-energies, residual power-law effects from the long-distance elastic contribution are still present even with an infinite-volume photon propagator and must be treated separately~\cite{Feng:2018qpx}; see section \ref{sec::long_distance}.
    
\end{itemize}

A program aimed at computing QED corrections to the muon $(g-2)$ based on this formalism has already started, with first results presented in~\cite{Parrino:2025afq,Erb:2025nxk}. In this paper, we instead apply the method to the computation of the charged–neutral pion mass difference $M_{\pi^+} - M_{\pi^0}$ at $\order{\alpha_\mathrm{em}}$ with the goal of demonstrating its validity and implementation in the case of a UV-finite observable that poses no issues related to renormalization. The present work therefore does not specifically illustrate the first bullet point listed above, but rather focuses on the aspects related to the finite-volume corrections, the continuum limit at fixed $\Lambda$ without counterterms, and the $\Lambda\to \infty$ extrapolation.  We do exploit to some extent the second bullet point, in that we opt for the combination of one local and one conserved vector current in our lattice implementation. In order to remove the $\Lambda$ regulator, we establish an interesting connection with phenomenology at fixed $\Lambda$ by comparing our lattice result with the elastic contribution to the pion mass splitting obtained from the Cottingham formula~\cite{Cottingham:1963zz}, which relates the electromagnetic correction to hadron masses to their self-energy.

Although the pion mass splitting at leading order in isospin-breaking effects is not a new observable and has already been determined in lattice QCD using different approaches~\cite{Frezzotti:2022dwn,Feng:2021zek}, the application of this novel method in this context provides valuable insights for future applications to other observables where the advantages described in the first two bullet points are more directly relevant, such as the electromagnetic corrections to the hadronic vacuum polarization contribution to the muon $(g-2)$ and the kaon mass splitting $M_{K^+}-M_{K^0}$. We also note that the second-order strong-isospin contribution to the pion mass splitting has recently been computed on the lattice~\cite{Bonanno:2025cmh}.

Historically, a dispersive sum rule was derived for the pion mass splitting in the chiral limit before QCD was established as the theory of the strong interaction~\cite{Das:1967it}. Detailed analyses~\cite{Donoghue:1993xb,OPAL:1998rrm,Davier:2005xq} showed that the experimentally measured vector and axial-vector spectral functions entering the sum rule were consistent with the pion mass splitting, within the sizable uncertainties. The sum rule has recently been generalized to include non-zero quark-mass effects~\cite{Bijnens:2026ohl}.

This work is organized as follows. In Section~\ref{sec::formalism} we introduce our formalism in which the leading-order electromagnetic correction is expressed using the PV-regulated photon propagator in position space. In Section~\ref{sec::setup} we provide details about the computational setup by making explicit the correlation functions needed for the calculation and fixing the notation. In this section we also illustrate the CLS lattice ensembles on which the data have been generated. Section~\ref{sec::long_distance}  deals with the infinite-time-separation limit and the reconstruction of the long-distance contribution via an analytical description of the elastic contribution. Finite-volume corrections are also discussed in this section. The results for the extrapolation to the physical point and to the continuum are detailed in Section~\ref{sec::chiral}. Finally, in Section~\ref{sec::pheno} we address the removal of the photon cutoff scale in the limit $\Lambda\to \infty$ and we make contact with the phenomenological description of the elastic contribution to the pion mass splitting using the Cottingham formula.

\section{Formalism \label{sec::formalism}}
Meson and baryon mass splittings can be determined by calculating the ground state contribution of two-point Euclidean correlation functions at zero momentum in the QCD+QED theory. 
In this paper, we follow the general approach of treating $\alpha_\mathrm{em}\neq0$ and $\delta m_\ell=m_d-m_u \neq 0$ as perturbative corrections to pure QCD with exact isospin symmetry, as pioneered in the lattice QCD context in~\cite{deDivitiis:2013xla}.
We then introduce the masses of the charged and neutral pions in QCD+QED as 
\begin{flalign}
M_\piplus &= M_\pi^\mathrm{isoQCD} + \delta [M_\piplus], \\ 
M_\pizero &= M_\pi^\mathrm{isoQCD} + \delta [M_\pizero],
\end{flalign}
where $\delta [M_\piplus]$ and $\delta [M_\pizero]$ account for electromagnetic and strong isospin breaking effects. The expansion of hadron masses around the isosymmetric point at first order gives corrections of order $\order{\alpha_\mathrm{em},\delta m_\ell}$. Due to isospin symmetry, the corrections of $\order{\delta m_\ell}$ to $M_\piplus$ and $M_\pizero$ cancel out when taking the difference, and therefore, the pion mass splitting at leading order is purely an electromagnetic effect,
\begin{flalign}
\Delta M_\pi = M_\piplus - M_\pizero = \delta [M_\piplus] - \delta [M_\pizero] = \order{\alpha_\mathrm{em}}.
\end{flalign}
The $\order{\alpha_\mathrm{em}}$ correction to a two-point correlation function amounts to inserting the photon propagator between the two quark lines and integrating over the two vertices. In this work we follow the same strategy and setup that are presented in the recent paper \cite{Erb:2025nxk}. The leading-order electromagnetic correction to the pion mass is expressed in the so-called \textit{mid-point} method, consisting in fixing the position of one of the two vector currents, integrating with respect to the position of the second one and taking the infinite-time separation limit of the interpolating operators,
\begin{flalign}\label{eq:midpoint}
\delta [M_{\pi^{+,0}}](\Lambda) =\frac{e^2}{2}\lim_{x_0\to \infty} \frac{\int \dd[4]z\; \delta^{\mu\nu}\mathcal{G}_\Lambda(z) \expval{O_{\pi^{+,0}}\big(\frac{x_0}{2}\big)J_\mu^\mathrm{em}(z)J_\nu^\mathrm{em}(0)O_{\pi^{+,0}}^\dagger\big(\frac{-x_0}{2}\big)}_\mathrm{isoQCD}}{\expval{O_{\pi^{+,0}}\big(\frac{x_0}{2}\big)O_{\pi^{+,0}}^\dagger\big(-\frac{x_0}{2}\big)}_\mathrm{isoQCD}}.
\end{flalign}
\begin{figure}
\centering
\includegraphics[width=0.48\linewidth]{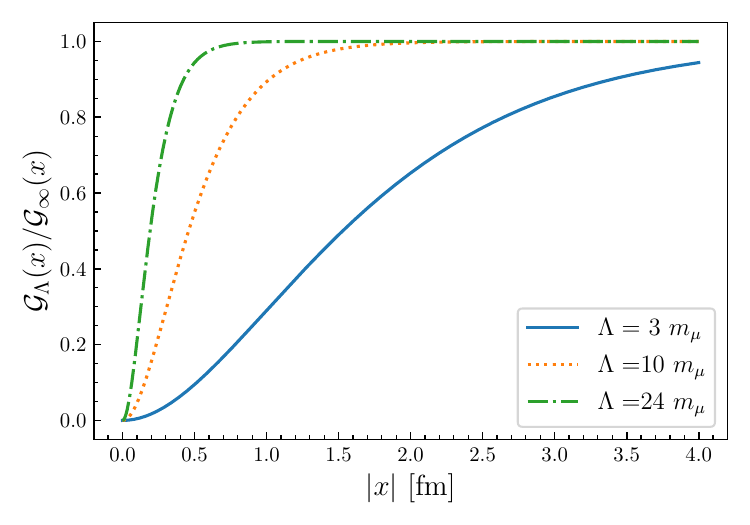}
\includegraphics[width=0.48\linewidth]{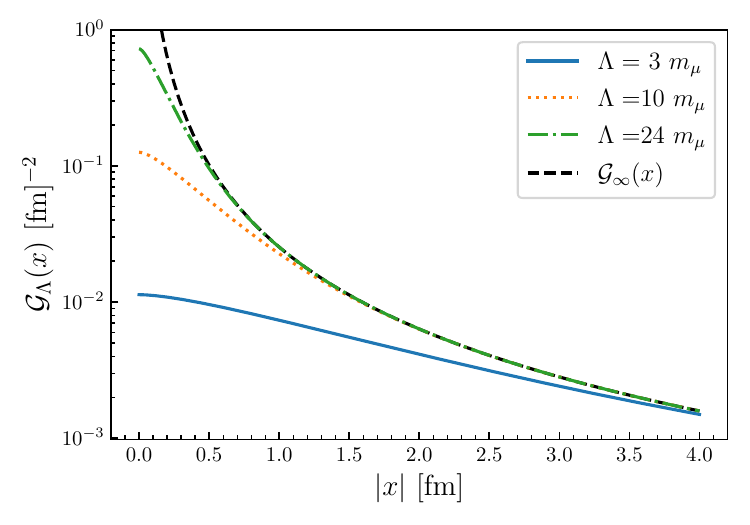}
\caption{\emph{Left plot:} ratio between the PV-regulated photon propagator and the unregulated one ($\Lambda=\infty$) for different values of the photon mass. \emph{Right plot:} comparison between the PV-regulated photon propagator and the unregulated one. }
\label{fig:photon_propagator}
\end{figure}
In the previous expressions $e^2=4\pi\alpha_\mathrm{em}=4\pi/137.036$ and $J_\mu^\mathrm{em}$ is the light-quark electromagnetic vector current, while $O_{\pi^{+}}(x)=\bar{d}\gamma_5 u(x)$ and $O_{\pi^0}=[\bar{u}\gamma_5 u(x)-\bar{d}\gamma_5 d(x)]/\sqrt{2}$ are the interpolating operators for the charged and neutral pion, respectively. The subscript isoQCD indicates that these expectation values are measured with respect to the gauge fields generated in a Monte Carlo simulation using the QCD action at the isosymmetric point.  The $x_0\to \infty$ limit projects the two external interpolating operators to a single-pion state at zero momentum, so that the mass correction can be conveniently rewritten as
\begin{flalign}
\delta [M_{\pi^{+,0}}](\Lambda) =\frac{e^2}{2}\int_{-\infty}^{+\infty} \dd{z_0}f_{\pi^{+,0}}(z_0,\Lambda),
\end{flalign}
with\footnote{Notice that compared to \cite{Erb:2025nxk} we do not include the factor $e^2/2$ in the definition of $f(z_0,\Lambda)$.}
\begin{flalign}\label{eq:f}
f_{\pi^{+,0}}(z_0,\Lambda)=\int \dd[3]{z} \mathcal{G}_\Lambda(z) \delta^{\mu\nu} \bra{\pi^{+,0},\bm{0}}\mathrm{T}\{J^\mathrm{em}_{\mu}(z)J^\mathrm{em}_{\nu}(0)\}\ket{\pi^{+,0},\bm{0}}.
\end{flalign}

In writing Eq.~\eqref{eq:midpoint} we have introduced the following  (doubly) Pauli-Villars-regulated photon propagator (PV-propagator) in Feynman gauge,
\begin{flalign}\label{eq:PV_propagator}
\mathcal{G}_\Lambda(x) = \frac{1}{4\pi^2 x^2} -2G_{\frac{\Lambda}{\sqrt{2}}}(x)+G_\Lambda(x),
\end{flalign}
where $G_m(x)=mK_1(m|x|)/(4\pi^2|x|)$ is the propagator of a massive scalar and $K_1(x)$ is the modified Bessel function of the second kind. This coordinate-space PV-propagator is  defined in infinite volume and the photon mass $\Lambda$ guarantees that $\mathcal{G}_\Lambda(x)$ is finite in the limit $|x|\to 0$.  The regulated propagator is compared to the unregulated one ($\Lambda = \infty$) in Fig.~\ref{fig:photon_propagator}. For convenience we express the photon cutoff as factors of the muon mass $m_\mu=0.10566~$GeV. The usage of the PV-propagator introduces a dependence of the pion mass correction on the scale $\Lambda$ and the physical result is obtained in the $\Lambda\to \infty$ limit. In order to keep the cutoff effects small, we should in general have $a\Lambda\ll 1$, where $a$ is the lattice spacing of the simulation. The pion mass splitting is a special observable in this sense because it does not require counterterms, namely, it is naturally UV-finite in the continuum even at $\Lambda = \infty$. By relying upon this feature we consider in this work photon cutoffs ranging from $3\,m_\mu$ up to the rather large value of $\Lambda=80 \, m_\mu\sim 8.5~$GeV. Such a large value of $\Lambda$ is in general not allowed for observables which require renormalization.

\section{Setup \label{sec::setup}}
\begin{figure}
\centering
\includegraphics[width=0.48\linewidth,valign=c]{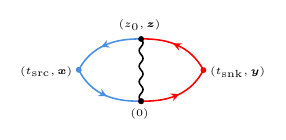}
\includegraphics[width=0.48\linewidth,valign=c]{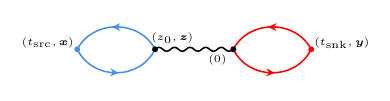}
\caption{The connected (left) and disconnected (right) diagrams relevant to the determination of $\Delta M_\pi$ at leading-order in $\alpha_\mathrm{em}$. See \cite{deDivitiis:2013xla} for the derivation.}
\label{fig:diagrams}
\end{figure}
On the lattice, the fermionic Wick contractions of the expectation value appearing in Eq.~\eqref{eq:midpoint} lead to two diagrams with different topologies, a quark-connected and a quark-disconnected diagram, shown in Fig.~\ref{fig:diagrams}. Up to the insertion of the photon propagator, these consist of the following four-point functions,
\begin{flalign}\label{eq:conn_minus_disco}
C_{\mu\nu}^\mathrm{conn}(z_0,t_\mathrm{sep},\bm{x},\bm{y},\bm{z})&=-\Tr\big[\gamma_\mu S(0,x)\gamma_5 S(x,z)\gamma_\nu S(z,y)\gamma_5 S(y,0)\big], \\[8pt]
\label{eq:disco_diagram}
C_{\mu\nu}^\mathrm{disc}(z_0,t_\mathrm{sep},\bm{x},\bm{y},\bm{z})&=\Tr\big[\gamma_\mu S(0,y)\gamma_5 S(y,0)]\times \Tr\big[\gamma_\nu S(z,x)\gamma_5 S(x,z)\big], 
\end{flalign}
where $S(u,v)$ is the fermion propagator for a degenerate light quark.  As derived in \cite{deDivitiis:2013xla}, the connected diagram appears both in $\delta M_\piplus$ and $\delta M_\pizero$ while the disconnected diagram only comes from $\delta M_\pizero$.  

We denote in bold character the spatial coordinates and, with reference to Fig.~\ref{fig:diagrams}, we have $x=(t_\mathrm{src},\bm{x})$, $y=(t_\mathrm{snk},\bm{y})$ and $z=(z_0,\bm{z})$. We call $t_\mathrm{sep}=|t_\mathrm{snk}-t_\mathrm{src}|$ the time separation between the two interpolating operators which, in isoQCD, corresponds to the pseudoscalar density $\bar\ell \gamma_5 \ell(x)$.  The expression for the pion mass splitting is thus
\begin{flalign}\label{eq:lattice_splitting}
\Delta M_\pi(\Lambda)= \frac{e^2}{2}(q_u-q_d)^2 \lim_{t_\mathrm{sep}\to \infty} \sum_{z_0=-\infty}^{+\infty} \frac{1}{{\sum_{\bm{x},\bm{y}}C^\mathrm{2pt}(t_\mathrm{sep},\bm{x},\bm{y})}}\sum_{\bm{x},\bm{y},\bm{z}} \delta^{\mu\nu}\mathcal{G}_\Lambda(z) 
\\[8pt]\nonumber
\times \big[C_{\mu\nu}^\mathrm{conn}(z_0,t_\mathrm{sep},\bm{x},\bm{y},\bm{z})+C_{\mu\nu}^\mathrm{disc}(z_0,t_\mathrm{sep},\bm{x},\bm{y},\bm{z})\big].
\end{flalign}
We denote by $q_u$ and $q_d$ the electric charges of the  up and down quarks in units of the electric charge of the positron. The object appearing in the denominator of Eq.~\eqref{eq:lattice_splitting} is the two-point function with the pseudoscalar interpolators evaluated at fixed time slice $t_\mathrm{sep}$,
\begin{flalign}
C^\mathrm{2pt}(t_\mathrm{sep},\bm{x},\bm{y})=\Tr\big[\gamma_5 S(x,y) \gamma_5S(y,x)\big].
\end{flalign}
In the mid-point method, the limit $t_\mathrm{sep}\to\infty$ projects the pion onto the ground state at zero momentum, ensuring that excited-state contamination is suppressed. In this setup, only one vertex of the photon propagator is summed over the volume, while the other is kept fixed. A sum over the spatial volume is performed for both the sink and source interpolators. This setup is computationally convenient, as it avoids the volume-squared sum over the fixed photon-propagator vertex. On the other hand, avoiding such a sum reduces the sampling of the corresponding vertex, thereby limiting the achievable statistical precision. 
\begin{table}[t]
\begin{center}
\footnotesize{
\begin{tabular}{ccccccc|c|c}
\toprule
ID & $L^3\times T$ & $a$ [fm] &$\beta$ & $L$ [fm] & $M_\pi$ [MeV] & $M_\pi L$ & $M_\mathrm{VMD}$ [MeV]  & $N_\mathrm{config}$ \\
\midrule
\midrule
H102 & $32^3\times 96$  & 0.085 & 3.40 & 2.7 & 358 & 4.9 & 856 & 1000\\
H105 & $32^3\times 96$  & 0.085 & 3.40 & 2.7 & 283 & 3.9 & 832 & 450\\
N101 & $48^3\times 128$ & 0.085 & 3.40 & 4.1 & 282 & 5.8 & 811 & 400 \\
C101 & $48^3\times 96$  & 0.085 & 3.40 & 4.1 & 222 & 4.6 & 784 & 2000\\
\midrule
B450 & $32^3\times 64$ & 0.075 & 3.46  & 2.4 & 422 & 5.1 & 902 & 200\\
S400 & $32^3\times 128$ & 0.075 & 3.46 & 2.4 & 355 & 4.3 & 843 & 2000\\
N452 & $48^3\times 128$ & 0.075 & 3.46 & 3.6 & 356 & 6.5 & 858 & 1000\\
N451 & $48^3\times 128$ & 0.075 & 3.46 & 3.6 & 291 & 5.3 & 832 & 1900\\
D450 & $64^3\times 128$ & 0.075 & 3.46 & 4.8 & 219 & 5.3 & 794 & 450\\
D452 & $64^3\times 128$ & 0.075 & 3.46 & 4.8 & 156 & 3.8 & 770 & 700\\
\midrule
N203 & $48^3\times 128$ & 0.064 & 3.55 & 3.0 & 349 & 5.4 & 861 & 1500 \\
N200 & $48^3\times 128$ & 0.064 & 3.55 & 3.0 & 286 & 4.4 & 831 & 800\\
D251 & $64^3\times 128$ & 0.064 & 3.55 & 4.1 & 286 & 5.9 & 830 & 500\\
D200 & $64^3\times 128$ & 0.064 & 3.55 & 4.1 & 202 & 4.2 & 767 & 600\\
\midrule
N302 & $48^3\times 128$ & 0.049 & 3.70 & 2.4 & 350 & 4.2 & 896 & 2000\\
J303 & $64^3\times 192$ & 0.049 & 3.70 & 3.1 & 260 & 4.1 & 834 & 600\\
\bottomrule
\bottomrule
\end{tabular}
}
\caption{\label{tab:ensembles} Parameters of CLS ensembles used in this work and approximate number of configurations.   The lattice spacing $a$ in physical units is based on the scale setting analyses of \cite{Strassberger:2021tsu,RQCD:2022xux}. The pion and the $M_\mathrm{VMD}$ masses are from \cite{Djukanovic:2024cmq}. The ensemble with ID B450 is not used in the calculation of the connected contribution.}
\end{center}
\end{table}

There are two important observations concerning the disconnected diagram. First, as pointed out in \cite{deDivitiis:2013xla}, the disconnected contribution vanishes in the $SU(2)$ chiral limit and, as such, its magnitude is $\mathcal{O}(\alpha_\mathrm{em}\cdot(m_u+m_d))$ rather than $\mathcal{O}(\alpha_\mathrm{em})$. Second, in the disconnected contribution there is no propagation of elastic intermediate states, but only inelastic ones, whose contribution is generally much smaller than that of the elastic states. For these two reasons, we argue that the disconnected contribution constitutes only a small correction to $\Delta M_\pi$. Therefore, in the remainder of the main text, we focus exclusively on the connected part. In Appendix~\ref{app:disco}, we provide numerical evidence supporting the smallness of the disconnected contribution through an explicit computation on a subset of our lattice ensembles. This estimate is then combined with the connected result at the end of Section~\ref{sec::pheno}.

In the rest of the paper, we then write the leading-order electromagnetic contribution to the pion mass splitting as
\begin{flalign}\label{eq:lattice_delta_mpi}
\Delta M_\pi(\Lambda) = \frac{e^2}{2}(q_u-q_d)^2 \lim_{t_\mathrm{sep}\to \infty} \sum_{z_0=-\infty}^{+\infty} f^\mathrm{latt}(z_0,\Lambda,t_\mathrm{sep}),
\end{flalign}
where we introduced the definition
\begin{flalign}\label{eq:flatt}
f^\mathrm{latt}(z_0,\Lambda,t_\mathrm{sep}) = \frac{\sum_{\bm{x},\bm{y},\bm{z}} \delta^{\mu\nu}\mathcal{G}_\Lambda(z)\big[C_{\mu\nu}^\mathrm{conn}(z_0,t_\mathrm{sep},\bm{x},\bm{y},\bm{z})\big]}{\sum_{\bm{x},\bm{y}}C^\mathrm{2pt}(t_\mathrm{sep},\bm{x},\bm{y})}.
\end{flalign}
We thus make explicit the dependence on the temporal separation between the two vector currents, given by the variable $z_0$. The connected four-point and two-point correlation functions have been computed on the fifteen lattice ensembles reported in Tab.~\ref{tab:ensembles}. They were generated within the Coordinated Lattice Simulations (CLS) effort with dynamical $O(a)$ improved Wilson-Clover up, down
and strange quarks and the tree-level $\order{a^2}$ improved L\"uscher-Weisz gauge action (see \cite{Bruno:2014jqa,Bali:2016umi,Djukanovic:2024cmq} and references therein for further information). This selection of ensembles includes four different lattice spacings in the range $a\in[0.049,0.085]$~fm and several heavier-than-physical pion masses in the range $M_\pi \in[160,360]~$MeV. In addition, there are pairs of ensembles with the same action parameters but different spatial extent that will be used in the next section to address the finite-size effects.

To compute the connected four-point function, we fix the temporal slice of the origin to $T/2$, where $T$ is the temporal extent of the lattice, and randomly generate  the spatial coordinate. The sink and source positions are placed symmetrically with respect to the origin, that is $t_\mathrm{snk}=T/2+t_\mathrm{sep}/2$ and $t_\mathrm{src}=T/2-t_\mathrm{sep}/2$. We compute a one-to-all propagator starting from the origin and then two sequential propagators over the sink and source vertices. 
In order to account for the wrap-around effects in the periodic boxes we correct the large time part of $C^\mathrm{2pt}$ by using a single-exponential fit.  In addition, we have generated data for several values of $t_\mathrm{sep}$ in order to address the $t_\mathrm{sep}\to \infty$ limit.

As for the electromagnetic currents, we employ the standard renormalized local definition,
\begin{flalign}\label{eq:Jc}
J_\mu^{\mathrm{em},l}(x)= \hat{Z}_V\bar\ell \gamma_\mu \ell(x),
\end{flalign}
for the current fixed at the origin (the renormalization constants have been taken from \cite{Gerardin:2018kpy}), and the point-split definition,
\begin{flalign}\label{eq:Jl}
J_\mu^{\mathrm{em},c}(x)=\frac{1}{2}\bigg[\bar{\ell}(x+\hat\mu)(\gamma_\mu+1)U^\dagger_\mu(x)\ell(x)\bigg]+\frac{1}{2}\bigg[\bar{\ell}(x)(\gamma_\mu-1)U_\mu(x)\ell(x+\hat{\mu})\bigg],
\end{flalign}
for the vertex that is integrated over. The point-split current, being exactly conserved on the lattice, does not require  renormalization factors. The local current at the origin is chosen for numerical convenience, since it requires a single point-to-all propagator, whereas a conserved current at the origin would require additional propagators. At fixed $\Lambda$, the PV-regulated propagator acts as a UV-finite weight function, so that any properly renormalized discretization of the currents leads to the same continuum limit. We have checked on a few ensembles that, at the present level of statistical errors, using the local current at both vertices gives results compatible with the local-conserved setup. Since the local-local choice has the same computational cost but requires an additional factor $Z_V$, we adopt the local-conserved setup throughout.

\section{Long-distance contribution and finite-volume effects\label{sec::long_distance}}

The discrete integration in Eq.~\eqref{eq:lattice_delta_mpi}  cannot in practice be extended up to $|z_0|=\infty$ since the number of available time slices is limited by $t_\mathrm{sep}$. On the other hand, as anticipated in the introduction and as  pointed out in \cite{Feng:2018qpx}, the photon propagator in infinite volume does not automatically remove the power-law suppressed finite-volume effects in the long-distance part of $f^\mathrm{latt}$. To tackle the problem, inspired by \cite{Feng:2018qpx,Feng:2021zek}, we separate the pion mass splitting into a short distance contribution, calculated on the lattice in the range $z_0\in[-t_c,+t_c]$, and a long distance contribution, which covers the remaining range $|z_0|\in[t_c,+\infty]$, calculated analytically directly in the continuum and infinite volume.  To derive the aforementioned analytical contribution, we work under the assumption that, at large time separations, the intermediate hadronic states between the two currents are elastic, namely only single-pion states dominate. We introduce relativistically normalized charged-pion states,
\begin{flalign}
\braket{\pi^+,\bm{p}}{\pi^+,\bm{p}'}=2E_{\bm{p}}L^3\delta_{\bm{p},\bm{p}'}\,,
\end{flalign}
and insert in Eq.~\eqref{eq:f} a finite-volume sum over all such single-pion states,
\begin{flalign}\label{eq:felast_tmp}
f^\mathrm{elast}(z_0,\Lambda,L)&= \int \dd[3]{z} \mathcal{G}_\Lambda(z)\delta^{\mu\nu}  \frac{1}{L^3}\sum_{\bm{p}} \frac{1}{4E_{\bm{p}}M_\pi} \\[8pt]
&\times \bra{\pi^+,\bm{0}} J_\mu^\mathrm{em}(z) \ket{\pi^+,\bm{p}}\bra{\pi^+,\bm{p}} J^\mathrm{em}_\nu(0) \ket{\pi^+,\bm{0}}. 
\end{flalign}
In the previous expression, we have fixed $z_0\ge 0$ and dropped the time-ordering operator. By translational invariance we have
\begin{flalign}\label{eq:invariance}
\bra{\pi^+,\bm{0}} J_\mu^\mathrm{em}(z) \ket{\pi^+,\bm{p}}=e^{(M_\pi-E_{\bm{p}})z_0+i\bm{p}\cdot \bm{z}}\bra{\pi^+,\bm{0}} J_\mu^\mathrm{em}(0) \ket{\pi^+,\bm{p}}.
\end{flalign}
The allowed momenta are discretized as $\bm{p}=\frac{2\pi}{L}\hat{n}$ and the energy is  $E_{\bm{p}}=\sqrt{M_\pi^2+\bm{p}^2}$. The matrix element in Eq.\eqref{eq:invariance} is expressed in terms of the electromagnetic pion form factor,
\begin{flalign}
\bra{\pi^+,\bm{p}'} J_\mu^\mathrm{em}(0)\ket{\pi^+,\bm{p}} = (p_\mu + p_\mu')F_\pi\big(-s\big),\qquad s = (p'-p)^2.
\end{flalign}
By inserting this decomposition, evaluated at $\bm{p}'=\bm{0}$, into Eq.~\eqref{eq:felast_tmp} we eventually obtain the elastic contribution in finite volume,
\begin{flalign}\label{eq:fpi_L}
f^\mathrm{elast}(z_0,\Lambda,L)= \frac{1}{2L^3}\int \dd[3]{z} \mathcal{G}_\Lambda(z) \sum_{\bm{p}} e^{(M_\pi-E_{\bm{p}})|z_0|+i\bm{p}\cdot\bm{z}} \cdot \frac{E_{\bm{p}}+M_\pi}{E_{\bm{p}}}\big[F_\pi(-s)\big]^2.
\end{flalign}

Note that we wrote this expression only for the charged pion. In fact, the elastic contribution of the electromagnetic correction to the neutral pion is simply zero since
\begin{flalign}
\bra{\pi^0,\bm{p}'} J_\mu^\mathrm{em}(0)\ket{\pi^0,\bm{p}} = 0
\end{flalign}
because of $C$ parity. Indeed, the electromagnetic correction to the neutral pion mass is much smaller compared to the charged one and, to a first approximation, would be $\Delta M_\pi \simeq \delta M_\piplus$ for $\Lambda$ not too large.
\begin{figure}
\centering
\includegraphics[width=0.48\linewidth]{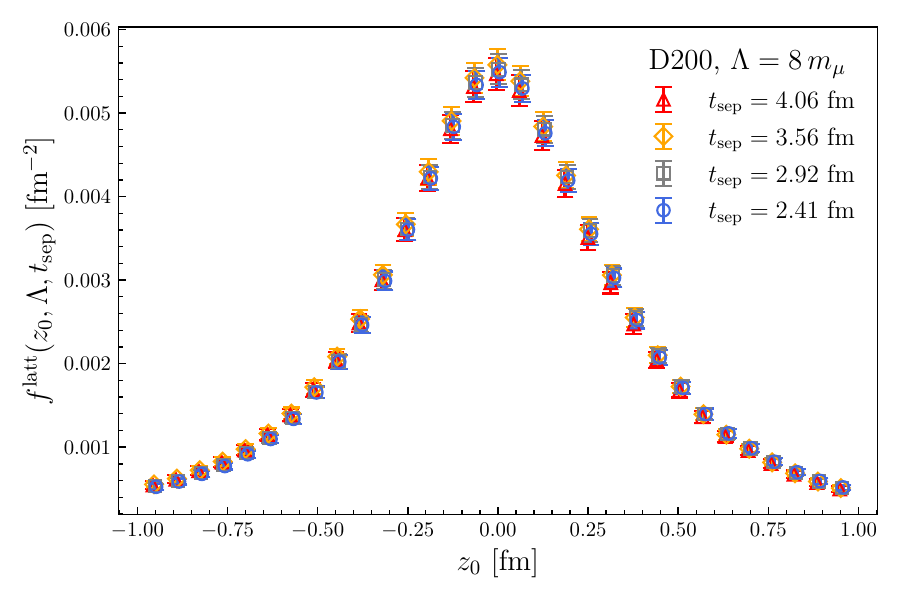}
\includegraphics[width=0.48\linewidth]{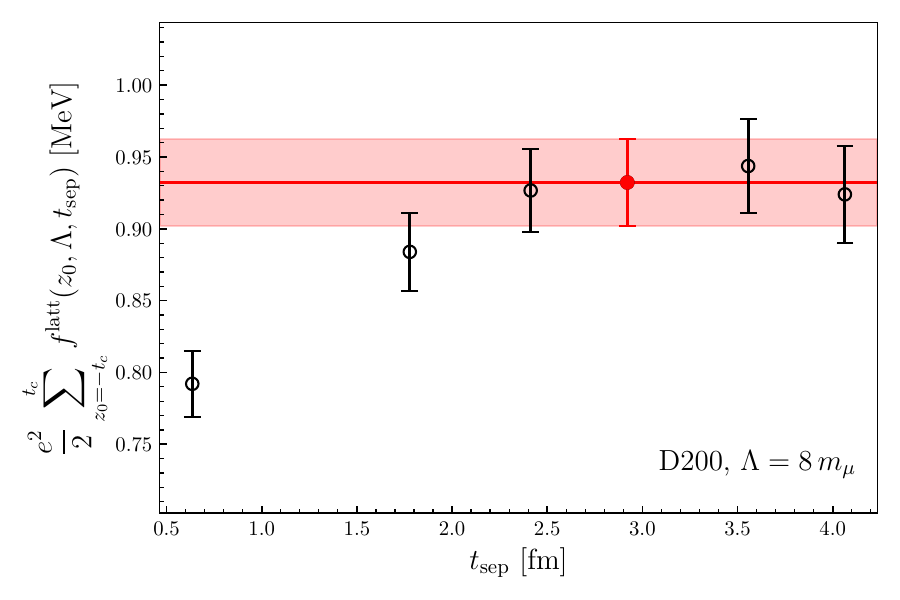}
\caption{\label{fig:tsep_limit} \emph{Left plot}: integrand $f^\mathrm{latt}$ for different values of $t_\mathrm{sep}$ in the case of the ensemble D200 at photon cutoff $\Lambda=8\,m_\mu$. The value of $t_c$ is set to 1~fm. The points are not yet corrected for the finite-volume effects and are shifted by a small amount with respect to the $z_0$ variable in order to improve the comparison. \emph{Right plot}: result of the integration of $f^\mathrm{latt}$ in the range $z_0\in[-t_c,+t_c]$ for several values of $t_\mathrm{sep}$ including the four values shown in the left plot. The data correspond to the ensemble D200 at photon cutoff $\Lambda=8\,m_\mu$. The red point corresponds to the one we used to quote the $t_\mathrm{sep}\to\infty$ limit and it is chosen in such a way that all the points at higher time separation are statistically compatible with it. The horizontal red band is drawn in order to facilitate the comparison between the different points.}
\end{figure}
The elastic contribution of Eq.~\eqref{eq:fpi_L} can also be expressed in infinite volume and, incorporating the PV-regulated photon propagator, the expression is given by
\begin{flalign}\label{eq:fpi_inf}
f^\mathrm{elast}(z_0,\Lambda,\infty)= \frac{1}{4\pi^2}\int_{0}^{\infty} \dd{|\bm{p}|} |\bm{p}|^2 e^{ |z_0|(M_\pi-E_{\bm{p}})}\cdot\frac{E_{\bm{p}}+M_\pi}{E_{\bm{p}}}\big[F_\pi(-s)\big]^2\\[8pt]\nonumber
\times \bigg[\frac{e^{-|\bm{p}||z_0|}}{2|\bm{p}|} - \frac{e^{-|z_0|\sqrt{|\bm{p}|^2+\Lambda^2/2}}}{\sqrt{|\bm{p}|^2+\Lambda^2/2}} + \frac{1}{2}\frac{e^{-|z_0|\sqrt{|\bm{p}|^2+\Lambda^2}}}{\sqrt{|\bm{p}|^2+\Lambda^2}}\bigg].
\end{flalign}
In order to evaluate $f^\mathrm{elast}$ numerically both in finite and infinite volume a phenomenological parametrization for the electromagnetic pion form factor must be provided. In this work we employ the Vector Meson Dominance (VMD) parametrization,
\begin{flalign}\label{eq:VMD}
F_\pi(-s)=\frac{1}{1+s/M_\mathrm{VMD}^2}.
\end{flalign}
The values of $M_\mathrm{VMD}$ used in the following analysis are listed in Table~\ref{tab:ensembles}. They have been previously obtained in the context of the leading-order hadronic vacuum polarization contribution to the muon $g-2$(see \cite{Djukanovic:2024cmq}). 

With this strategy, the pion mass splitting is obtained according to
\begin{flalign}\label{eq:splitting_full}
\Delta M_\pi (\Lambda) = \frac{e^2}{2} \bigg[ \lim_{t_\mathrm{sep}\to \infty}\sum_{z_0=-t_c}^{+t_c} f^\mathrm{latt}(z_0,\Lambda,t_\mathrm{sep}) +  2 \int_{t_c}^{\infty} \dd{z_0} f^\mathrm{elast}(z_0,\Lambda,\infty)\bigg]. 
\end{flalign}

Notice that $f^\mathrm{elast}$ is a symmetric function with respect to $z_0=0$, from which we get the factor 2 in front of the integral for $f^\mathrm{elast}$. The time $t_c$ is chosen in such a way that  $f^\mathrm{latt}(z_0,\Lambda)$ is statistically compatible with $f^\mathrm{elast}(z_0,\Lambda,L)$ and for our ensembles, this typically occurs for $t_c\in [1,1.2]$~fm.  For each ensemble we compute the integral of $f^\mathrm{latt}(z_0,\Lambda)$ in the range $z_0\in[-t_c,+t_c]$ for different temporal separations and perform the $t_\mathrm{sep}\to\infty$ limit through a plateau analysis, that is by looking for a saturation with respect to $t_\mathrm{sep}$ within the statistical noise. An example of this procedure, repeated for all the ensembles and for all the values of the photon cutoff, is shown in Figure~\ref{fig:tsep_limit} for the ensemble D200 at $\Lambda=8\,m_\mu$. We observe that typically the plateau is reached for $t_\mathrm{sep} >2.5$~fm in our ensembles.
\begin{figure}
\centering
\includegraphics[width=0.48\linewidth]{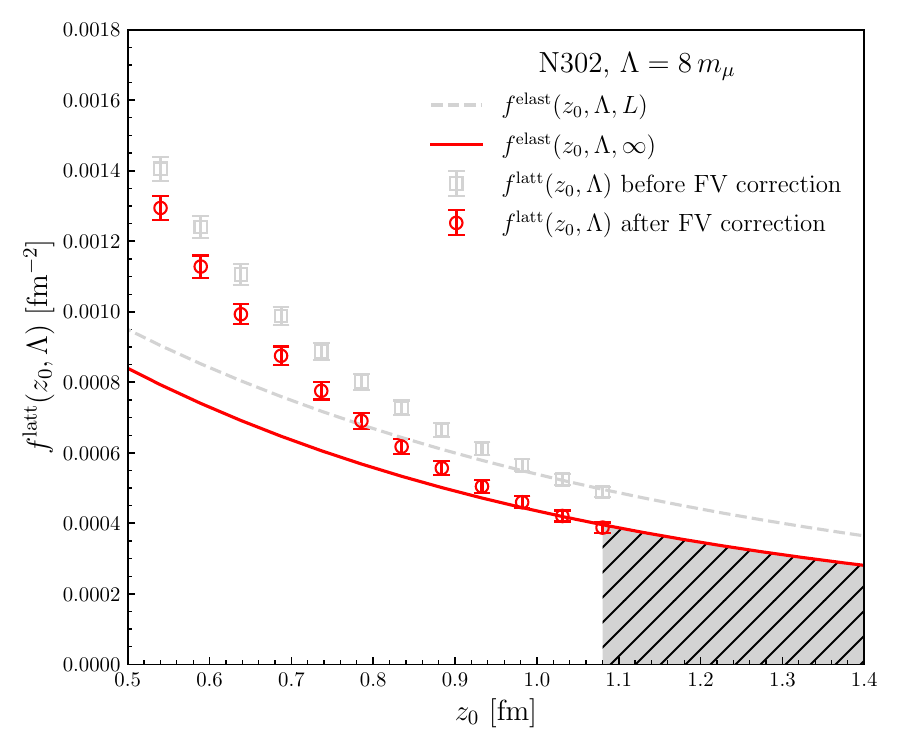}
\includegraphics[width=0.48\linewidth]{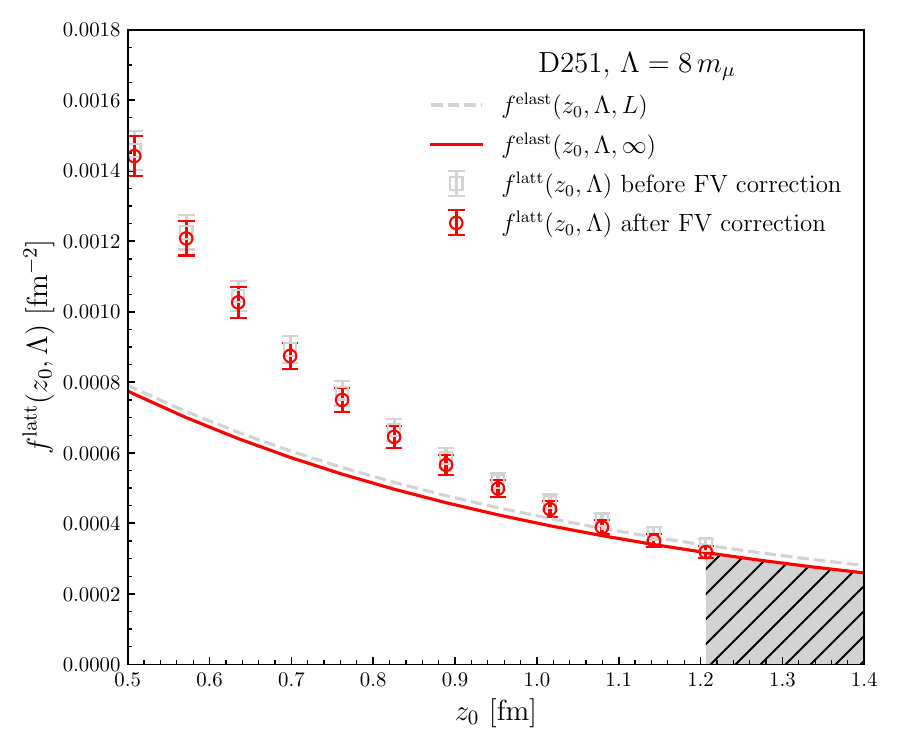}
\caption{\label{fig:fve}\emph{Left plot}: reconstruction of the long-distance contribution to $f^\mathrm{latt}(z_0,\Lambda)$ and correction for finite-volume effects for the ensemble N302 ($L=2.4$~fm) at $\Lambda=8m_\mu$. The gray squares represent the original lattice data (already in the $t_\mathrm{sep}\to \infty$ limit), while the red circles correspond to the lattice data after applying the finite-volume correction as in Eq.~\eqref{eq:fve_correction}. The dashed gray line and the red solid line represent $f^\mathrm{elast}(z_0,\Lambda)$ in finite and infinite volume, respectively. The shaded area indicates the region where $f^\mathrm{elast}(z_0,\Lambda)$, already in infinite volume, replaces $f^\mathrm{latt}(z_0,\Lambda)$. For this ensemble, we set $t_c=1.1$~fm. \emph{Right plot}: same as the left plot, but for the ensemble D251 ($L=4.1$~fm) at $\Lambda=8m_\mu$. In this case, we set $t_c=1.2$~fm.}
\end{figure}

Since the elastic contribution contains information on the lightest states propagating between the two photon vertices, it is expected to be much more sensitive to finite-volume effects than the inelastic contribution. We can thus employ it to correct the lattice data in the range $[-t_c,+t_c]$ for finite-volume effects. In particular, we correct the data according to
\begin{flalign}\label{eq:fve_correction}
f^\mathrm{latt}(z_0,\Lambda)\big|_{L=\infty} =f^\mathrm{latt}(z_0,\Lambda)\big|_{L}+ \big[f^\mathrm{elast}(z_0,\Lambda,\infty)-f^\mathrm{elast}(z_0,\Lambda,L)\big] ,
\end{flalign}
where $f^\mathrm{latt}$ on the right-hand side is already taken in the $t_\mathrm{sep}\to \infty$ limit. The effect of this correction is illustrated in the two panels of Figure~\ref{fig:fve} for the N302 and D251 ensembles at photon cutoff $\Lambda=8\,m_\mu$. For the former, which, with $L=2.4$~fm, is among the ensembles with the smallest volumes, the correction is sizeable (although still small compared to the result after integration over $z_0$), whereas it is much smaller for the latter, where the spatial extent of the box is $L=4.1$~fm.

\begin{figure}
\centering
\includegraphics[width=\linewidth]{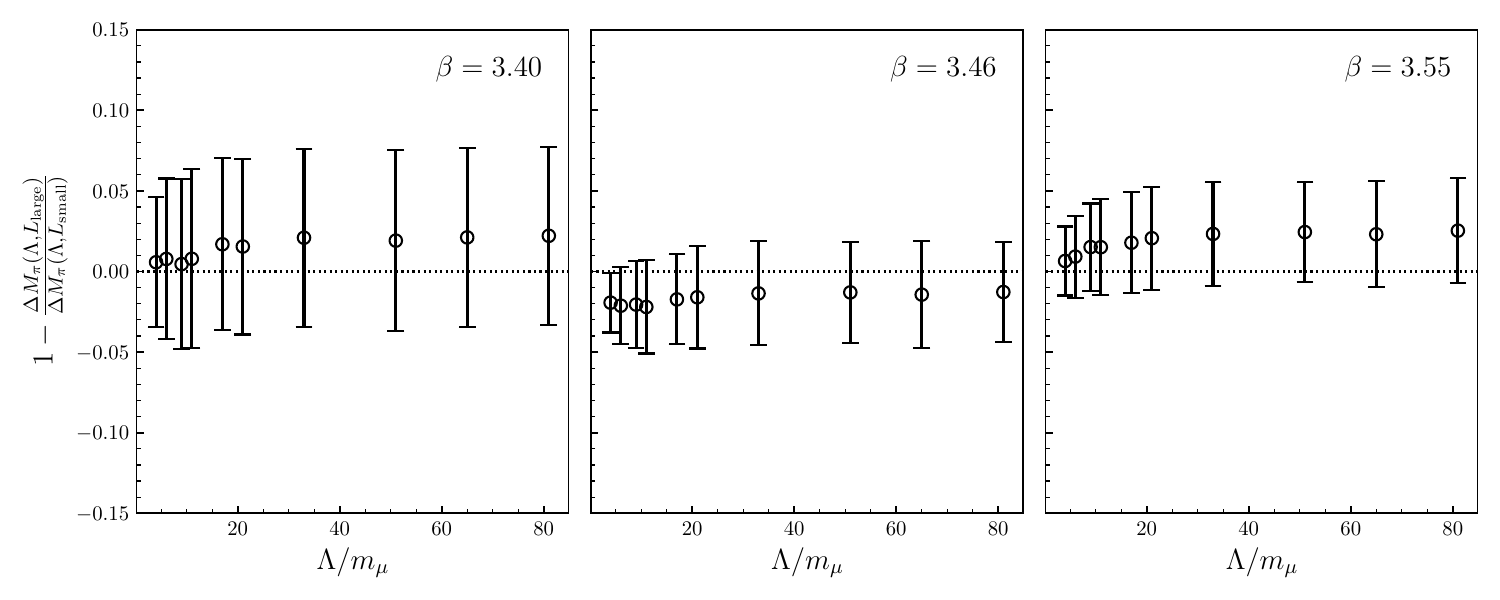}
\caption{\label{fig:fve_pull} Relative deviation in $\Delta M_\pi(\Lambda)$ between the large- and small-volume lattice data after correcting for the finite-volume effects for all the values of $\Lambda$. The three plots refer to the pairs N101-H105 (left), N452-S400 (middle) and D251-N200 (right). The two ensembles in each pair only differ by the temporal $T$ and spatial extent $L$.}

\end{figure}

To assess possible residual finite-volume effects arising from the inelastic contribution, we explicitly compute the relative deviation of $\Delta M_\pi(\Lambda)$, after applying the correction in Eq.~\eqref{eq:fve_correction}, between a large and a smaller-volume ensemble at fixed action parameters. Using the ensembles listed in Table~\ref{tab:ensembles}, this quantity can be evaluated at three different values of $\beta$ and for pion masses of approximately $280$~MeV and $350$~MeV. The results, shown in Figure~\ref{fig:fve_pull}, indicate that the relative deviation is compatible with zero in all cases.
Therefore, after computing the pion mass splitting according to Eq.~\eqref{eq:splitting_full}, with $f^\mathrm{latt}$ corrected as in Eq.~\eqref{eq:fve_correction}, we treat our results as effectively corresponding to the infinite-volume limit and do not consider additional finite-volume corrections.

The values of the splitting for each value of $\Lambda$ and for each individual ensemble, after correcting for finite-volume effects, are reported in Table~\ref{tab:all_numbers} of Appendix~\ref{app:material}.

\section{Extrapolation to the physical point\label{sec::chiral}}
In order to make contact with phenomenology and experiment we extrapolate our results to the physical point $M_{\pi,\mathrm{phys}}=134.98~$MeV and to the continuum, $a=0$. For convenience, we consider the squared mass splitting which is directly related to $\Delta M_\pi$ as 
\begin{flalign}\label{eq:conversion}
\big[M_\piplus^2-M_\pizero^2\big](\Lambda)=2M_\pi\Delta M_\pi(\Lambda) + \order{\alpha^2_\mathrm{em}}.
\end{flalign}
Inspired by ChPT  (see for instance \cite{Hayakawa:2008an,Gasser:2003hk}) and by noticing that the pion mass splitting at leading order in isospin-breaking effects is purely due to electromagnetism, namely there is no dependence on the difference of light quark masses, we  consider the following ansatz to model the dependence on the lattice spacing and on the pion mass,
\begin{flalign}\label{eq:fit_ansatz}
\big[M_\piplus^2-M_\pizero^2\big](\Lambda,a,M_\pi)=\frac{e^2}{2} \bigg[c_0 +c_a a^2  + c_\pi^{(1)} M_\pi^2 + c_\pi^{(2)} M_\pi^2 \log\bigg(\frac{M_\pi^2}{\mu^2}\bigg)+c_{a\pi}(aM_\pi)^2\bigg].
\end{flalign}
The ansatz contains five parameters, which are determined by fitting to the lattice data. The coefficient $c_0$ represents a constant term, while $c_a$ is proportional to $a^2$. The coefficients $c_\pi^{(1)}$ and $c_\pi^{(2)}$ encode, respectively, a linear and a logarithmic dependence on $M_\pi^2$, and $c_{a\pi}$ accounts for the interplay between the lattice spacing and the pion mass. The scale $\mu$ is fixed at $4\pi f_\pi$ with $f_\pi=90$~MeV. The ansatz contains no term linear in $a$. In isoQCD, hadron masses computed with an $\mathcal{O}(a)$-improved action are automatically free of $\mathcal{O}(a)$ effects since, unlike generic matrix elements, they do not require the improvement of any operator. At $\mathcal{O}(\alpha_{\rm em})$, however, the photon couples to the quarks through the lattice vector currents, and a complete $\mathcal{O}(a)$ improvement of the mass correction would require improved currents, the counterpart of an electromagnetic clover term in the action. Since the currents of Eqs.~\eqref{eq:Jc}-\eqref{eq:Jl} are not $\mathcal{O}(a)$-improved, a residual $\mathcal{O}(a\alpha_{\rm em})$ effect must in general be expected, though it need not be dominant in our range of lattice spacings. We have thus repeated the extrapolation replacing the $a^2$ term with a term linear in $a$, and found results statistically compatible with those obtained with the ansatz of Eq.~\eqref{eq:fit_ansatz}. At the present level of precision, the data do not discriminate between the two forms, and we adopt Eq.~\eqref{eq:fit_ansatz} in the following.

The dependence on $\Lambda$ is treated in a separate step; accordingly, the extrapolation to the physical point is performed independently for each value of $\Lambda$. To assign a reliable systematic uncertainty to the extrapolated results, we perform a series of fits in which the number of unconstrained parameters is varied. In particular, we consider three strategies.

In the first, any dependence on the lattice spacing is neglected by setting $c_a = c_{a\pi} = 0$, yielding three free parameters. In the second, only a linear dependence on $a^2$ is assumed by setting $c_{a\pi} = 0$, with a total of four free parameters. Finally, in the third, no constraints are imposed on $c_a$ and $c_{a\pi}$. This corresponds to the fit with the largest number of free parameters. The coefficients $c_\pi^{(1)}$ and $c_\pi^{(2)}$ are always included and treated as free parameters. To combine the results from the different fits we perform a weighted average \cite{Jay:2020jkz} inspired by the Akaike Information Criterion (AIC) \cite{Akaike:1974vps}. We assign to each fit a weight (in the form proposed in \cite{Borsanyi:2020mff}) defined by
\begin{flalign}\label{eq:weights}
w_k\propto \exp\big[-(\chi^2_k +2N_\mathrm{params}^k-N_\mathrm{points}^k)/2\big].
\end{flalign}
The weights of the fits entering the average are normalized in such a way that they sum up to one. As an estimator for the central value we consider the weighted average
\begin{flalign}
x = \sum_{k=1}^{N} w_k x_k,
\end{flalign}
where $N$ is the total number of fits and $x_k$ is the central value of each individual fit. As the systematic error, we compute the weighted average of the spread of each fit with respect to $x$,
\begin{flalign} \label{eq:systematic_error}
\sigma_\mathrm{syst}^2 = \sum_{k=1}^{N} w_k(x_k-x)^2,
\end{flalign}
and quote as total error the combination in quadrature with the statistical uncertainty.
\begin{figure}[t]
\centering
\includegraphics[width=0.93\linewidth]{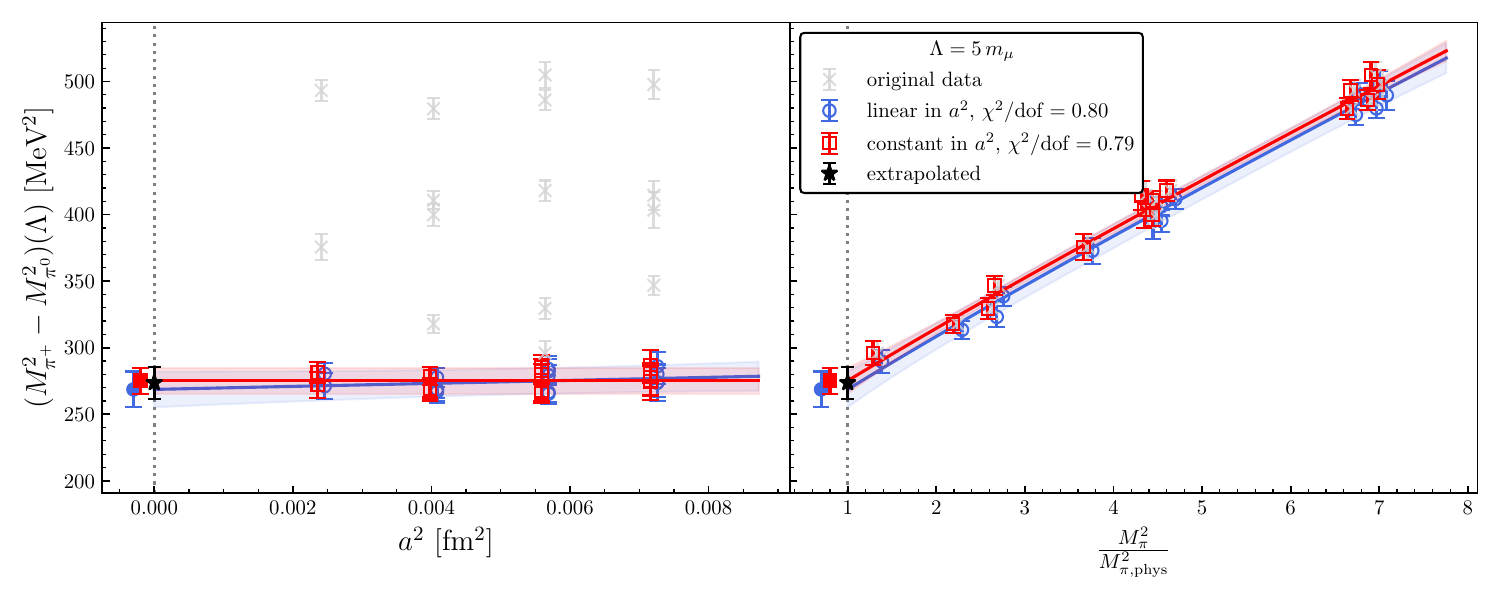}
\includegraphics[width=0.93\linewidth]{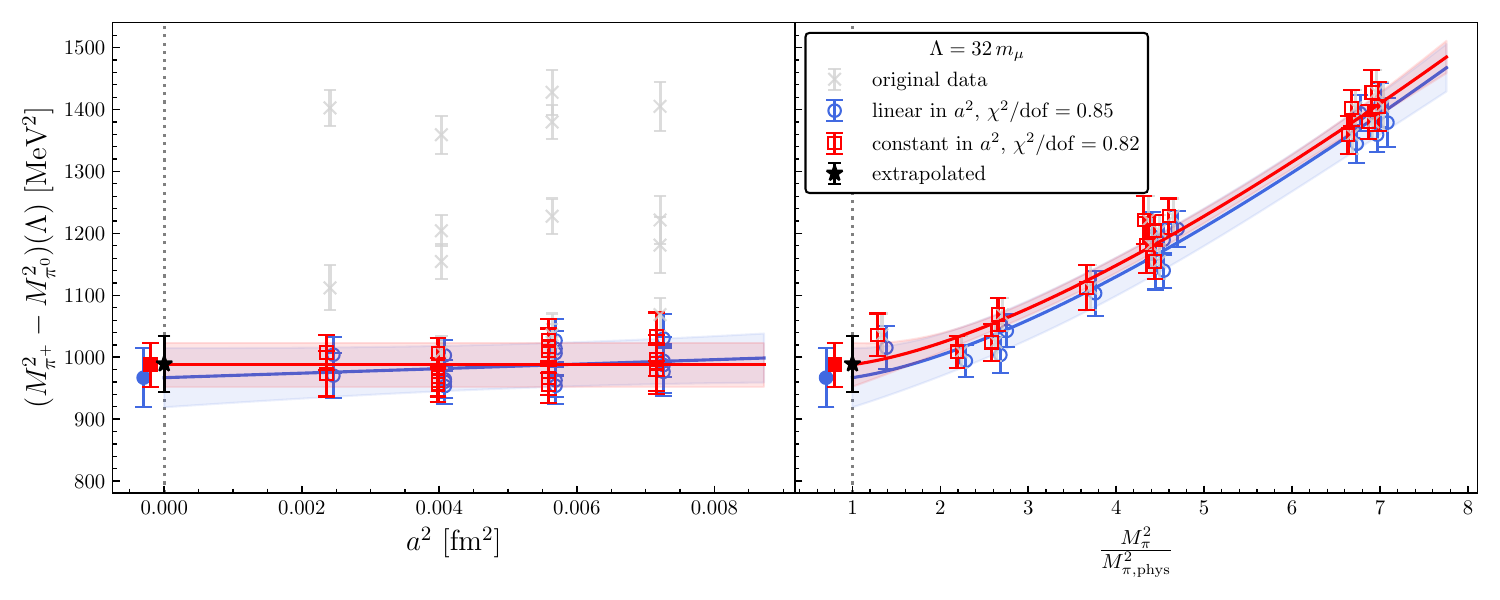}
\includegraphics[width=0.93\linewidth]{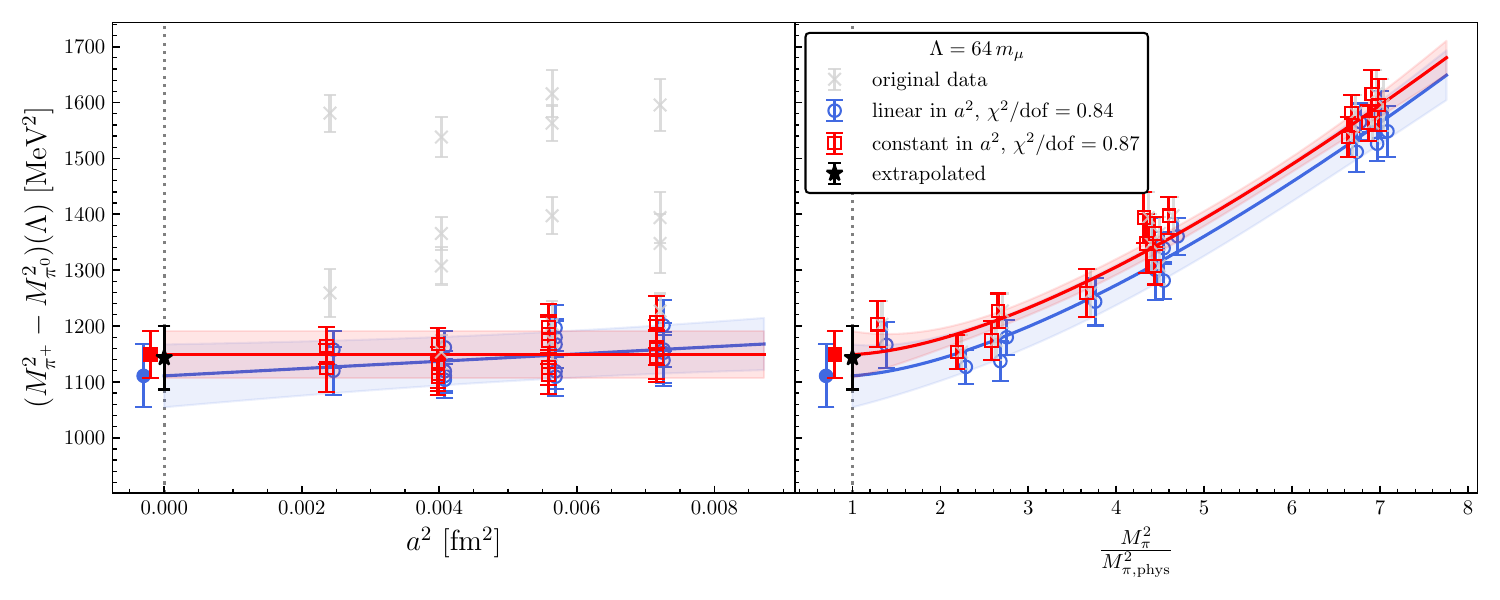}
\caption{\label{fig:chiral_cont}  Chiral and continuum extrapolation of the squared pion mass splitting for the photon cutoff $\Lambda=5\,m_\mu$ (top panel), $\Lambda=32\,m_\mu$ (middle panel) and $\Lambda=64\,m_\mu$ (bottom panel). In each panel we show on the left the  $a^2$-dependence and on the right the normalized $M_\pi^2$-dependence.  The gray crosses correspond to the original lattice data, while the other points are those corrected by using the fit ansatz, imposing $M_\pi=M_\pi^\mathrm{phys}$ to isolate the $a^2$-dependence, and imposing $a=0$ to isolate the $M_\pi^2$-dependence, respectively. The blue circles are the results assuming a linear dependence on $a^2$ while the red squares assume no dependence on the lattice spacing. The result of each individual fit is represented by the filled points. The black star is the result of the extrapolation (already including the systematic error) obtained from the weighted average of 19 different fits. }
\end{figure}
\begin{figure}
\centering
\includegraphics[width=0.7\linewidth]{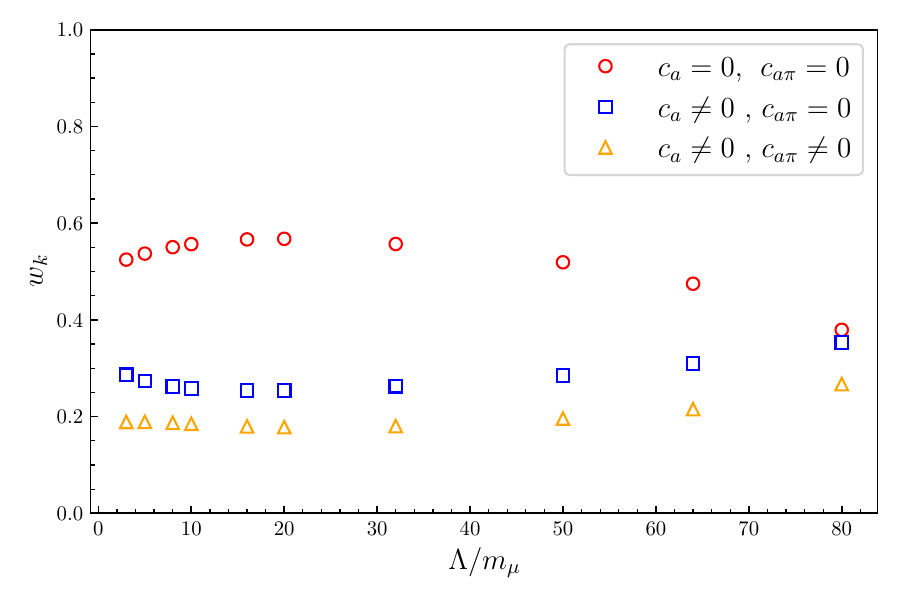}
\caption{\label{fig:weights} Weights of the fits as function of the cutoff scale $\Lambda$ determined using all the ensembles. The weights are normalized to sum up to one. Red circles correspond to a fit constant in $a^2$, blue squares to a fit linear in $a^2$ and orange triangles to a fit with a simultaneous dependence on $a^2$ and $a^2M_\pi^2$.}
\end{figure}
Figure~\ref{fig:chiral_cont} shows the results of the extrapolation for the first two fit strategies at $\Lambda = 5 m_\mu, 32 m_\mu,$ and $64 m_\mu$, including all fifteen data points. As can be seen, both the fit with no dependence on $a^2$ and the fit with a linear dependence on $a^2$ provide a good description of the data, with a reduced $\chi^2$ below 1 in all cases. The two extrapolations provide results that are fully compatible within statistical uncertainties. Indeed, in the second fit strategy, the coefficient $c_a$ is always determined with a relative uncertainty of order 100\% or larger, indicating a very mild dependence on the lattice spacing.

However, it is important to investigate this behavior as a function of the photon cutoff $\Lambda$. The PV regulator directly influences the cutoff effects associated with the photon. In our setup, the kernel $\mathcal{G}_\Lambda(z)$ is evaluated on the lattice sites and summed over the position $z$ of one of the two currents. As shown in Fig.~\ref{fig:photon_propagator}, the regulated propagator coincides with the unregulated one at distances $|z|\gg 1/\Lambda$, while it is strongly suppressed for $|z|\lesssim 1/\Lambda$ and remains finite at the origin, $\mathcal{G}_\Lambda(0)=\Lambda^2\ln 2/(16\pi^2)$. For $a\Lambda\ll 1$, the kernel varies slowly on the scale of the lattice spacing, so that the lattice sum provides an accurate approximation of the continuum integral, and the region $|z|\sim a$, where the lattice correlation function is most affected by discretization effects, gives only a small contribution. The cutoff effects associated with the photon are therefore governed by $a\Lambda$, in addition to those of isoQCD. What we find empirically is a mild dependence on the lattice spacing for $a\Lambda<1$, and a more significant one for $a\Lambda\gtrsim 1$, when the kernel is no longer resolved by the lattice. 

To quantify this effect, Figure~\ref{fig:weights} shows the normalized weights of Eq.~\eqref{eq:weights} for the three fit strategies as a function of $\Lambda$, obtained by including all ensembles. The fit with no dependence on $a^2$ (red circles) is dominant, with a weight that is roughly constant and above 0.5 up to $\Lambda = 20\, m_\mu$. Beyond this point, its contribution decreases in favor of the fit including a linear term in $a^2$ (blue squares), and the two become equally important at the largest photon cutoff considered in this work, $\Lambda = 80\, m_\mu$. This transition is consistent with the fact that, for all four lattice spacings used in this work (see Table~\ref{tab:ensembles}), $a\Lambda < 1$ for $\Lambda \le 20\, m_\mu$, while $a\Lambda > 1$ for $\Lambda \ge 32\, m_\mu$. The situation is expected to be more pronounced for observables that are not naturally UV-finite (e.g. the kaon mass splitting, $\Delta M_K = M_{K^+} - M_{K^0}$). Finally, the fit including the additional parameter $c_{a\pi}$ (orange triangles) is subdominant compared to the other two, indicating that, at the current level of statistical precision, no significant dependence on an $a^2 M_\pi^2$ term is observed. Consistently, the coefficient $c_{a\pi}$ is always determined with an uncertainty of order 100\%. Without the regulator, i.e.\ with the propagator $1/(4\pi^2 z^2)$, the kernel is singular at the origin and largest precisely in the region $|z|\sim a$. Therefore, a prescription for the point $z=0$ is required, and the cutoff effects are no longer controlled by the small parameter $a\Lambda$.  For the UV-finite pion mass splitting, the continuum limit still exists, but its approach is less well controlled. This is illustrated by the analysis of App.~\ref{app:material}, where the $\Lambda\to\infty$ extrapolation is performed at fixed lattice spacing, which approximates the use of the unregulated propagator. In this case, a fit including a cutoff-effect term is preferred and the total uncertainty is about 50\% larger.

Finally, to test the stability of our results, we consider several additional fits to the three fits discussed above. These are obtained by considering only the first two fit strategies and: i) removing all points corresponding to the coarsest lattice spacing; ii) removing all points with $M_\pi > 300$~MeV; and iii) removing individual points that may be affected by statistical fluctuations. In this way, we obtain a total of 19 fits, all with a reduced $\chi^2$ below 1.2, which are then combined according to the procedure illustrated above.

The final results for the squared pion mass difference  including statistical, systematic, and total uncertainties, are reported in Table~\ref{tab:results}. The correlation matrix between the extrapolated points corresponding to different values of $\Lambda$ is
\begin{flalign}\label{eq:correlation}
\rho(\Lambda_1,\Lambda_2)=
\begin{pmatrix}
1.00 & 0.99 & 0.97 & 0.96 & 0.95 & 0.94 & 0.94 & 0.93 & 0.93 & 0.93 \\
0.99 & 1.00 & 1.00 & 0.99 & 0.98 & 0.98 & 0.98 & 0.98 & 0.98 & 0.97 \\
0.97 & 1.00 & 1.00 & 1.00 & 1.00 & 1.00 & 0.99 & 0.99 & 0.99 & 0.99 \\
0.96 & 0.99 & 1.00 & 1.00 & 1.00 & 1.00 & 1.00 & 1.00 & 0.99 & 0.99 \\
0.95 & 0.98 & 1.00 & 1.00 & 1.00 & 1.00 & 1.00 & 1.00 & 1.00 & 0.99 \\
0.94 & 0.98 & 1.00 & 1.00 & 1.00 & 1.00 & 1.00 & 1.00 & 1.00 & 0.99 \\
0.94 & 0.98 & 0.99 & 1.00 & 1.00 & 1.00 & 1.00 & 1.00 & 1.00 & 1.00 \\
0.93 & 0.98 & 0.99 & 1.00 & 1.00 & 1.00 & 1.00 & 1.00 & 1.00 & 1.00 \\
0.93 & 0.98 & 0.99 & 0.99 & 1.00 & 1.00 & 1.00 & 1.00 & 1.00 & 1.00 \\
0.93 & 0.97 & 0.99 & 0.99 & 0.99 & 0.99 & 1.00 & 1.00 & 1.00 & 1.00 \\
\end{pmatrix},
\end{flalign}
which shows a high correlation.
\begin{table}[t]
\begin{center}
\footnotesize{
\begin{tabular}{cc|c}
\toprule
$\Lambda/m_\mu$ & $\big[M_\piplus^2-M_\pizero^2\big](\Lambda)$~MeV$^2$  & $\big[M_\piplus^2-M_\pizero^2\big]^\mathrm{elast}(\Lambda)$~MeV$^2$\\
\midrule
\midrule
  3 &  154(5)(3)[ 6] &  157 \\
  5 &  273(10)( 6)[12] &  270 \\
  8 &  436(16)(10)[19] &  420 \\
 10 &  528(19)(12)[23] &  503 \\
 16 &  731(27)(18)[32] &  686 \\
 20 &  824(30)(20)[36] &  770 \\
 32 &  989(37)(25)[45] &  920 \\
 50 & 1100(42)(30)[52] & 1023 \\
 64 & 1143(45)(34)[56] & 1063 \\
 80 & 1174(47)(42)[63] & 1092 \\
\bottomrule
\bottomrule
\end{tabular}
}
\caption{\label{tab:results}  Our results for the squared mass splitting after extrapolation to the continuum limit and physical point for different values of $\Lambda$. The numbers in the first and second round brackets correspond to statistical and systematic error, respectively. The numbers in the square brackets are the total errors obtained as the combination in quadrature of the statistical and systematic ones. The elastic contribution is determined using the Cottingham formula as explained in Section~\ref{sec::pheno}.}
\end{center}
\end{table}

\section{Final results and comparison with phenomenology \label{sec::pheno}}
The final step in determining the electromagnetic pion mass splitting is the removal of the cutoff scale $\Lambda$. In this context, a phenomenologically relevant aspect is the separation between the elastic contribution, where only pions appear as intermediate states, and the inelastic one, which includes all other intermediate states. Beyond its phenomenological interest, this separation is particularly convenient for handling the $\Lambda \to \infty$ limit.

The pion mass splitting plays a special role in this respect, as it is ultraviolet finite. By contrast, for a generic hadron, the mass splitting becomes divergent in the limit $\Lambda \to \infty$ if the appropriate counterterms are not included. Since the elastic contribution is finite, the divergence originates entirely from the inelastic part. Moreover, at moderate $\Lambda$ the elastic contribution typically provides the dominant effect, which further motivates a separation between elastic and inelastic components.

The lattice data obtained in the previous section naturally contain both contributions. The elastic contribution can be determined phenomenologically based on the Cottingham formula \cite{Cottingham:1963zz} which expresses the electromagnetic self-energy in terms of the forward Compton tensor. In particular, we follow \cite{Stamen:2022uqh}  and write the electromagnetic elastic correction to the charged pion square mass as
\begin{flalign}\label{eq:cottingham}
\delta[M_\piplus^2]^\mathrm{elast} = \frac{ie^2}{2}\int \frac{\dd[4]{k}}{(2\pi)^4}\frac{g^{\mu\nu}T_{\mu\nu}^\mathrm{elast}(k)}{k^2+i\varepsilon},
\end{flalign}
where $T_{\mu\nu}^\mathrm{elast}$ is the elastic Compton tensor in the forward direction,
\begin{flalign}\label{eq:compton_tensor}
T_{\mu\nu}^\mathrm{elast}(k)=i\Bigg[\frac{(2p+k)_\mu(2p+k)_\nu}{(p+k)^2-M_\pi^2}+\frac{(2p-k)_\mu(2p-k)_\nu}{(p-k)^2-M_\pi^2}-2g_{\mu\nu}\Bigg]\big[F_\pi(k^2)\big]^2.
\end{flalign}
In the previous equation $p$ is the on-shell momentum of the pion. The contraction of the tensor yields
\begin{flalign}
ig^{\mu\nu}T_{\mu\nu}^\mathrm{elast}(k)=\frac{2k^2(3k^2-4M_\pi^2)-16(p\cdot k)^2}{(k^2)^2-4(p\cdot k)^2}\big[F_\pi(k^2)\big]^2.
\end{flalign}
We now substitute the contracted tensor into Eq.~\eqref{eq:cottingham}  and modify the photon propagator in order to include the Pauli-Villars mass $\Lambda$ according to Eq.~\eqref{eq:PV_propagator}. The correction reads
\begin{flalign}\label{eq:elastic_contribution}
\delta[M_\piplus^2]^\mathrm{elast}(\Lambda) =\frac{\alpha_\mathrm{em}}{8\pi}\int_{0}^{\infty}\dd{s} s\big[F_\pi(-s)]^2  \bigg[4W(s)+\frac{s}{M_\pi^2}\big(W(s)-1\big)\bigg] \hat{\mathcal{G}}_\Lambda(s),
\end{flalign}
where $W(s)=\sqrt{1+4M_\pi^2/s}$ and $s=-k^2$ which is obtained after Wick rotating $k_0$ to the imaginary axis. $\hat{\mathcal{G}}_\Lambda(s)$ is the PV-regulated propagator in momentum space,
\begin{flalign}\label{eq:Gmomentum}
\hat{\mathcal{G}}_\Lambda(s)= \Delta_0(s) -2 \Delta_{\frac{\Lambda}{\sqrt{2}}}(s)+\Delta_\Lambda(s),
\end{flalign}
with $\Delta_m(s)=1/(s+m^2)$. For the electromagnetic form factor we use again the parametrization of Eq.~\eqref{eq:VMD} with $M_\mathrm{VMD}=M_\rho=770$~MeV. Notice that the above definition of the elastic contribution differs from the one used in Section~\ref{sec::long_distance}. There we assumed propagation of single-pion intermediate states and external zero-momentum pions (see Eq.~\eqref{eq:felast_tmp})    to model the long-time part of the connected contribution Eq.~\eqref{eq:flatt}. The Cottingham formula given in Eq.~\eqref{eq:cottingham}, together with the elastic Compton tensor Eq.~\eqref{eq:compton_tensor}, is instead a self-energy diagram in which the pion carries non-vanishing spatial momentum and propagates both in the $s$-channel and $u$-channel.

As observed in Section~\ref{sec::long_distance}, the neutral pion receives no correction from the elastic states and, therefore,
\begin{flalign}
[M_\piplus^2-M_\pizero^2]^\mathrm{elast}(\Lambda)=\delta[M_\piplus^2]^\mathrm{elast}(\Lambda).
\end{flalign}

Eq.~\eqref{eq:elastic_contribution} can be evaluated numerically for the same values of $\Lambda$ as those for which we have computed the squared pion mass splitting. The results are reported in Table~\ref{tab:results}. We convert $M_{\pi^+}^2 - M_{\pi^0}^2$ and $\delta[M_\piplus^2]^\mathrm{elast}$ to $\Delta M_\pi$ and $\Delta M_\pi^\mathrm{elast}$ by dividing by $2 M_\pi^\mathrm{phys}$, which according to Eq.~\eqref{eq:conversion} is valid up to $\mathcal{O}(\alpha_\mathrm{em}^2)$. 
\begin{figure}
\centering
\includegraphics[width=0.48\linewidth]{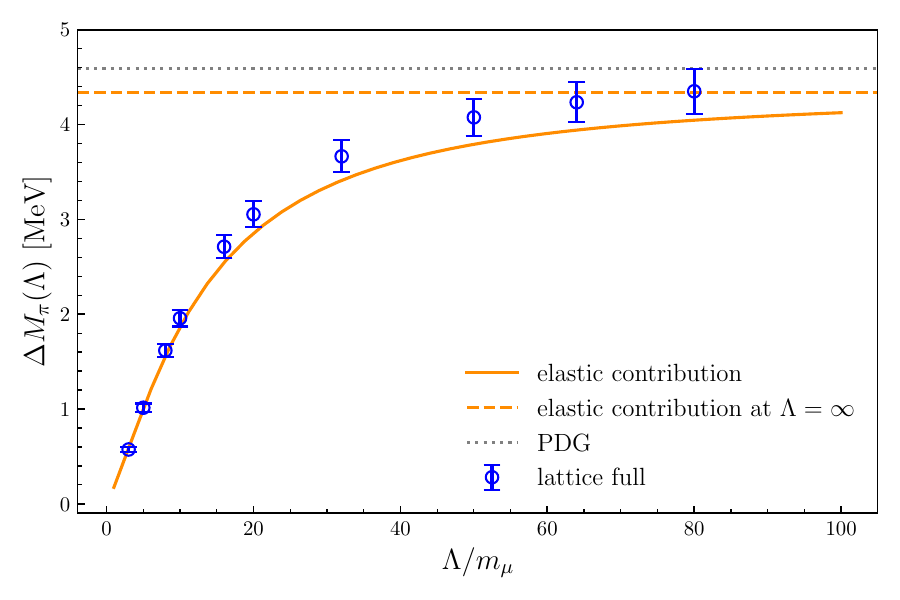}
\includegraphics[width=0.48\linewidth]{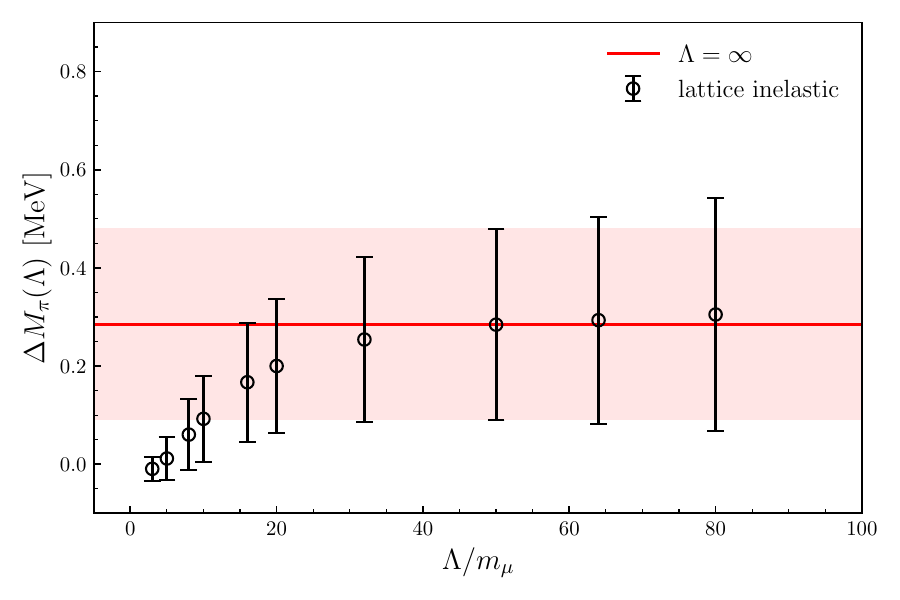}
\caption{\emph{Left}: our lattice determination for $\Delta M_\pi(\Lambda)$ (blue circles) as a function of $\Lambda$ compared to the elastic contribution (solid line) determined from the forward Compton amplitude. The error bars already include the systematic error determined from the extrapolation to the physical point and to the continuum. The dashed line corresponds to the elastic contribution at $\Lambda = \infty$ and the dotted line corresponds to the experimental value for the pion mass splitting.  \emph{Right}: dependence of the inelastic contribution on $\Lambda$ and $\Lambda \to \infty$ extrapolation (horizontal band). }
\label{fig:phenomenology}
\end{figure}
The lattice results for $\Delta M_\pi(\Lambda)$ are compared to the elastic contribution obtained from the Cottingham formula in the left plot of Figure~\ref{fig:phenomenology}. In the same plot, we also show the experimental value $\Delta M_\pi^\mathrm{pdg} = 4.5936(5)~\text{MeV}$ \cite{ParticleDataGroup:2024cfk}, as well as the elastic contribution evaluated at $\Lambda = \infty$, which reads
\begin{flalign}\label{eq:elast_inf}
\Delta M_\pi^\mathrm{elast}(\Lambda=\infty)=4.33~\text{MeV}.
\end{flalign}

As can be seen, the elastic term accounts for more than 90\% of the mass splitting for all values of $\Lambda$; therefore, almost the entire $\Lambda$ dependence is driven by the elastic contribution. The growth of $\Delta M_\pi(\Lambda)$ with increasing $\Lambda$ is rather slow and is not yet saturated at the largest value considered, $\Lambda = 80\,m_\mu$.

To perform the $\Lambda \to \infty$ extrapolation, we separate the full lattice result into elastic and inelastic contributions. We thus define the inelastic pion mass splitting as
\begin{flalign}
\Delta M_\pi^\mathrm{inelast}(\Lambda)=\Delta M_\pi(\Lambda)-\Delta M_\pi^\mathrm{elast}(\Lambda).
\end{flalign}
The resulting points are shown in the right plot of Figure~\ref{fig:phenomenology}, and the precision is sufficient to determine the inelastic contribution with statistical significance. The saturation with respect to $\Lambda$ for $\Delta M_\pi^\mathrm{inelast}$ is faster than for the elastic contribution, and the onset of a plateau within uncertainties is observed around $\Lambda = 20\,m_\mu$, for which $a\Lambda <1$.

To determine the inelastic contribution in the $\Lambda \to \infty$ limit, we perform a constant weighted average using the four points corresponding to the largest values of $\Lambda$ ($32$, $50$, $64$, and $80$ times $m_\mu$). The result of this procedure, represented by the red band in the aforementioned figure, is
\begin{flalign}
\Delta M_\pi^\mathrm{inelast} = 0.29(16)(12)[20]\;\text{MeV},
\end{flalign}
where the uncertainties in round brackets correspond to the statistical and systematic errors, respectively. Adding the elastic contribution of Eq.~\eqref{eq:elast_inf} we finally obtain the following determination for the connected part,
\begin{flalign}\label{eq:final_conn}
\Delta M_\pi = 4.62(16)(12)[20]\;\text{MeV}.
\end{flalign}
The addition of the disconnected contribution determined in Appendix~\ref{app:disco} gives
\begin{flalign}
M_\piplus - M_\pizero=\Delta M_\pi+\Delta M_\pi^\mathrm{disc} = 4.56(19)(12)[22]\;\text{MeV}.
\end{flalign}
Our determination for the electromagnetic pion mass splitting is in very good agreement with $\Delta M_\pi^\mathrm{pdg}$ and with two lattice calculations not based on a PV-regulated propagator, $4.622(95)$~MeV \cite{Frezzotti:2022dwn} and $4.534(60)$~MeV \cite{Feng:2021zek}, both including the disconnected contribution.

\section{Conclusion\label{sec::conclusion}}
In this work, we have performed a lattice QCD calculation of the mass difference between the charged and neutral pion induced by electromagnetic effects. At order $\alpha_\mathrm{em}$, the mass splitting receives two contributions, a connected and a disconnected one. We have generated data for both contributions using several CLS ensembles. For the connected part, which accounts for more than 99\% of the total splitting, we have investigated finite-volume effects, the continuum limit, and the extrapolation to the physical point. For the disconnected part, we have only studied the extrapolation to the physical point. Our final result reads
\begin{flalign}
M_{\pi^+} - M_{\pi^0} = 4.56(19)(12)[22]~\text{MeV},
\end{flalign}
which is in excellent agreement with the experimental value and with other lattice QCD determinations based on different strategies.

To implement the photon on the lattice, we have employed the formalism introduced in \cite{Biloshytskyi:2022ets}, which makes use of a position-space Pauli–Villars (PV)–regulated photon propagator defined in infinite volume. The use of an infinite-volume propagator, together with a modelling of the long-distance elastic contribution, allows for a controlled treatment of the infinite-volume limit, leaving only exponentially suppressed finite-volume effects in the lattice data, which we further correct using the elastic contribution. The PV-regulated photon propagator also introduces an additional scale, $\Lambda$, which must be removed in the $\Lambda \to \infty$ limit. The pion mass splitting at order $\alpha_\mathrm{em}$ is UV finite in the $a \to 0$ limit and, exploiting this property, we have considered values of $\Lambda$ as large as $80\,m_\mu = 8.5$~GeV.

To study the $\Lambda \to \infty$ limit, we have made use of the phenomenological prediction for electromagnetic mass corrections to mesons provided by the Cottingham formula, suitably modified to incorporate our regulated propagator. This approach allows for a useful separation of the pion mass splitting into elastic and inelastic components. Most of the $\Lambda$ dependence is carried by the elastic contribution, which can be treated analytically, while the inelastic component exhibits a mild residual dependence on $\Lambda$ and is determined directly from our data. This observation is particularly relevant for observables that are not UV-finite unless properly renormalized, for which the condition $a\Lambda \ll 1$ must be satisfied.

The strategy presented in this work demonstrates that implementing the photon directly in infinite volume with a Pauli–Villars regulator is viable and well controlled. This formalism is currently being applied to other observables. In particular, \cite{Erb:2025nxk} marks the beginning of a program aimed at computing electromagnetic corrections to the hadronic vacuum polarization (HVP) contribution to the muon $(g-2)$. In addition to employing the PV-regulated photon propagator, this calculation makes use of the covariant coordinate-space (CCS) representation \cite{Meyer:2017hjv} for the HVP.

Another quantity currently under investigation within this framework is the kaon mass splitting, $M_{K^+} - M_{K^0}$, induced by electromagnetic effects. 
It provides the relevant counterterm not only for the proton–neutron mass splitting, but also for the isospin-violating part of the vacuum polarization contribution to the muon $(g-2)$.

The statistical precision of the calculation presented in this work is limited by the numerical setup adopted, in which the sum over the spatial volume is performed only for one of the two internal vertices connected by the photon propagator. Although this setup is computationally less demanding, it leads to a reduction of the statistical sampling. In this respect, the recent work \cite{Erb:2026dtl} explores alternative setups in which the dependence on the two internal vertices is factorized, allowing both spatial sums to be computed. This investigation, carried out in the context of electromagnetic corrections to the muon $(g-2)$, yields very promising results towards making this formalism not only practically viable but also competitive in precision with more conventional approaches. Such a setup would also improve the determination of the disconnected contribution, whose uncertainty is currently not negligible in our total error.
\acknowledgments 

 Calculations for this project were performed on the HPC cluster “Mogon II” at JGU Mainz. We are grateful to our colleagues in the CLS initiative for sharing ensembles.
We acknowledge the support of Deutsche Forschungsgemeinschaft (DFG, German Research Foundation) through the research unit FOR\;5327 “Photon-photon interactions in the Standard Model and beyond” (project No.~458854507), and through the Cluster of Excellence “Precision Physics, Fundamental Interactions and Structure of Matter” (PRISMA+ EXC 2118/1) funded within the German Excellence Strategy (project ID 390831469).
 \appendix
 \section{Disconnected contribution to $\Delta M_\pi$ \label{app:disco}}

The disconnected diagram which contributes to the pion mass splitting at order $\order{\alpha_\mathrm{em}}$ requires a different strategy to be evaluated compared to the connected one, which is discussed in the main text. From Eq.~\eqref{eq:lattice_splitting} and Eq.~\eqref{eq:disco_diagram} we see that the computation of the following Wick-contraction is needed,
\begin{flalign}\label{eq:disco}
\sum_{\bm{x},\bm{y},\bm{z}} \delta^{\mu\nu}\mathcal{G}_\Lambda(z) &C^\mathrm{disc}_{\mu\nu}(z_0,t_\mathrm{sep},\bm{x},\bm{y},\bm{z})\\ \nonumber
&=\sum_{\bm{x},\bm{y},\bm{z}} \delta^{\mu\nu}\mathcal{G}_\Lambda(z)\Tr\big[\gamma_\mu S(0,y)\gamma_5 S(y,0)]\times \Tr\big[\gamma_\nu S(z,x)\gamma_5 S(x,z)\big] ,
\end{flalign}
corresponding to the diagram depicted on the right of Figure~\ref{fig:diagrams}. All the propagators refer to a light mass-degenerate quark. The first trace can  easily be evaluated by computing the one-to-all propagator $S(0,y)$  using a point source in the origin and then summing over the vertex $y$ at fixed time-slice $y_0=t_\mathrm{snk}$. The propagator $S(y,0)$ is obtained from the first one by exploiting the $\gamma_5--$hermiticity property. The use of a point source at the origin for this trace is dictated by the mid-point method adopted throughout this work. Keeping one vertex of the photon propagator fixed reduces the evaluation of the kernel $\mathcal{G}_\Lambda$ to a single volume sum. Stochastic sources for both traces would allow both vertices to be summed over, at the price of a double volume sum over $\mathcal{G}_\Lambda(z-w)$. Efficient strategies for this case, based on a factorization of the photon propagator, are explored in Ref.~\cite{Erb:2026dtl} but not adopted here where the goal is just to obtain an estimate for the disconnected diagram. The second trace instead entails a spatial sum over both the vertices of the propagator for which the computation of the all-to-all propagator  $S(z,x)$ at fixed time slice $x_0=t_\mathrm{src}$ is required. Again, $\gamma_5-$hermiticity is used to obtain $S(x,z)$. We employ stochastic wall sources \cite{Boyle:2008rh} and estimate the propagator according to 
\begin{flalign}\label{eq:stochastic_estimator}
S(z,x)\simeq \frac{1}{N_\mathrm{hits}} \sum_{n=1}^{N_\mathrm{hits}} \phi^{(n)}(z) \otimes \eta^{(n)\dagger}(x),
\end{flalign}
where $\phi^{(n)}(z)$ is the solution of the system $D\phi^{(n)}=\eta^{(n)}$ and $\eta^{(n)}$ is a wall source defined as
\begin{flalign}
\eta^{(n)}_{t_\mathrm{src}}(x)^\alpha_c = 
\left \{
\begin{aligned}
&\eta^{(n)}(\bm{x})^\alpha_c, && \text{if}\; x_0=t_\mathrm{src} \\
	&0, && \text{otherwise}.
\end{aligned} \right.
\end{flalign}
In the previous relations, $D$ is the Dirac operator while $\alpha$ and $c$ are respectively spin and colour indices. For the stochastic source, any set of numbers satisfying 
\begin{flalign}
\lim_{N_\mathrm{hits}\to \infty} \frac{1}{N_\mathrm{hits}}\sum_{n=1}^{N_\mathrm{hits}} \eta^{(n)}(\bm{x})^\alpha_c \eta^{(n)\dagger}(\bm{y})^{\beta}_d=\delta_{\bm{x}\bm{y}}\delta_{\alpha\beta}\delta_{cd}
\end{flalign}
\begin{figure}
\centering
\includegraphics[width=\linewidth]{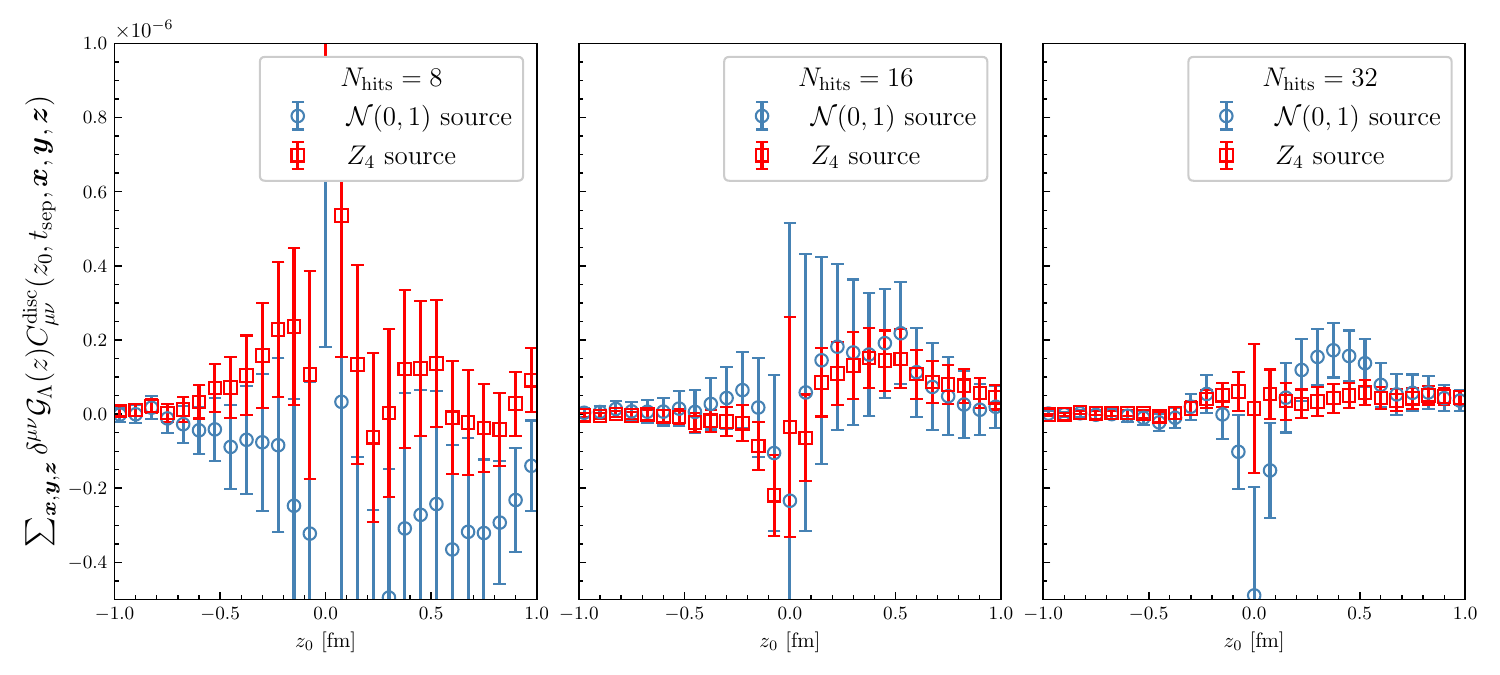}
\caption{\label{fig:sources_comparison} Comparison between the disconnected diagram generated using normally-distributed sources (blue points) and $Z_4$ sources (red squares) for $N_\mathrm{hits}=8$ (left plot), 16 (central plot) and 32 (right plot). The data correspond to the ensemble B450 and are generated using the same $N_U=200$ configurations.  The data correspond to $\Lambda=16\,m_\mu$. }
\end{figure}
is in principle valid for the estimator in Eq.~\eqref{eq:stochastic_estimator} but some choices lead to a reduced variance of the estimator.  As a first step, we make use of both  $\mathcal{N}(0,1)$ sources, which are obtained by sampling a normal distribution with vanishing center and unit variance, and $Z_4$ sources~\cite{Dong:1993pk}, which are obtained by sampling with uniform probability within the set $Z_4=\{1,-1,i,-i\}$. In both cases, we generate undiluted stochastic sources. The comparison between the two different sources is done by computing the diagram of Eq.~\eqref{eq:disco} on the ensemble B450 (see Table~\ref{tab:ensembles}) for $N_U=200$ gauge configurations and for $N_\mathrm{hits}=8$, $16$ and $32$.  We fix the temporal separation between source and sink to $t_\mathrm{sep}=2.5~$fm and compute the contribution for the same values of the scale $\Lambda$ that are used for the connected one.  The comparison is shown in Figure~\ref{fig:sources_comparison} where it can be appreciated that the usage of $Z_4$ sources allows for a much less noisy determination of the disconnected contribution. We have not further investigated the interplay between stochastic and gauge noise, which is beyond the scope of this work, and generated the disconnected diagrams for other ensembles using $Z_4$ sources and $N_\mathrm{hits}=32$. In particular, in addition to the B450 ensemble we generated data for the ensembles D450, D251 and N451 by fixing, in all the cases, $t_\mathrm{sep}=3~$fm which, according to the analysis performed on the connected contribution, is enough to isolate the correction to the ground state. Given the smallness of the disconnected contribution, we compute the splitting by summing over $z_0$ in the range $[-t_\mathrm{sep}/2,+t_\mathrm{sep}/2]$ without applying finite-volume corrections which, in this case, would require a data-driven approach since the elastic contribution used to parametrize the long-distance part in the connected case does not apply here.
\begin{figure}
\centering
\includegraphics[width=0.48\linewidth]{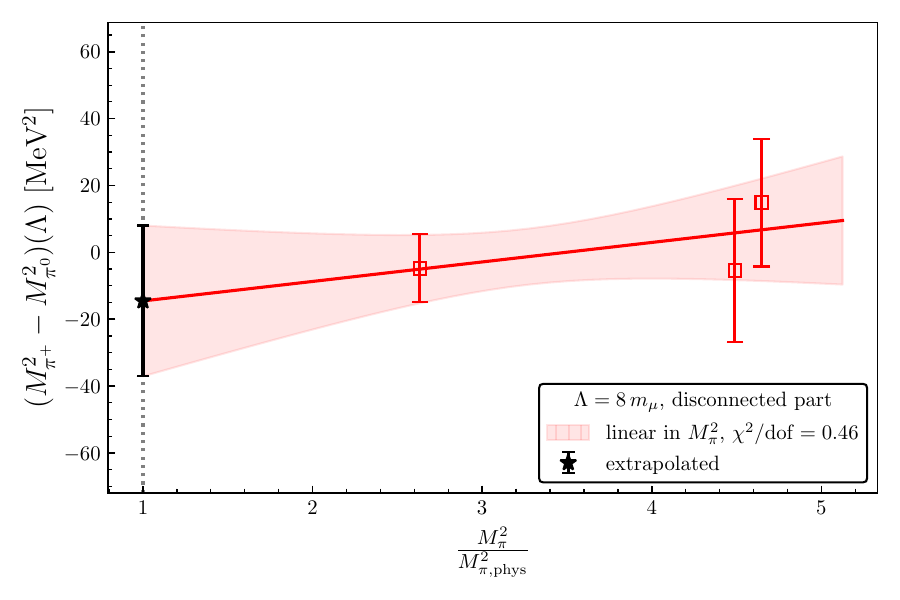}
\includegraphics[width=0.48\linewidth]{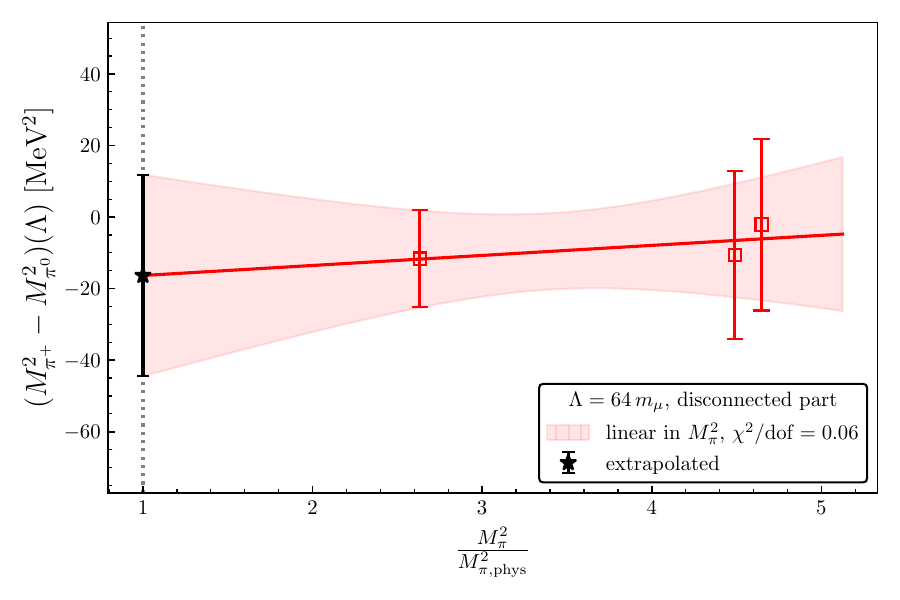}
\caption{\label{fig:extrapolation_disconnected} Extrapolation to the physical point for the disconnected contribution in case of $\Lambda=8\,m_\mu$ (left plot) and $\Lambda=64\,m_\mu$ (right plot). Due to the reduced number of points we only adopt one fit ansatz linear in $M_\pi^2$. The black star corresponds to the extrapolated point.}
\end{figure}

The extrapolation to the physical point is done looking at the squared pion mass difference obtained according to Eq.~\eqref{eq:conversion}. In the extrapolation we exclude the data corresponding to the B450 ensemble since it is at SU(3)-symmetric point and its pion mass is too far away from the physical point. Given the fact that we have data for only three independent points we account for the dependence on $M_\pi^2$ but not on $a^2$ which, relying on the findings in Section~\ref{sec::chiral}, is expected to play a negligible role. We extrapolate $[M^2_\piplus-M^2_\pizero](\Lambda)$ to the physical point using a simple two-parameter fit linear in $M_\pi^2$.  Examples of the extrapolation in case of $\Lambda=8\,m_\mu$ and $\Lambda=64\,m_\mu$ are shown in Figure~\ref{fig:extrapolation_disconnected}. As  can be seen, the assumption of an $M_\pi^2$-linear dependence is enough to extrapolate the data and the addition of further parameters, in combination with the fact that the disconnected diagrams are generally more noisy than connected contractions, would only lead to an overfitting of the data points. The results of the extrapolations, for all the values of $\Lambda$, are shown in Table~\ref{tab:results_disconnected}. 

\begin{table}[t]
\begin{center}
\footnotesize{
\begin{tabular}{cc}
\toprule
$\Lambda/m_\mu$ & $\big[M_\piplus^2-M_\pizero^2\big](\Lambda)$~MeV$^2$ (disconnected)\\
\midrule
\midrule
  3 & -10(15)  \\
  5 & -16(20)  \\
  8 & -15(23)  \\
 10 & -17(24)  \\
 16 & -13(28)  \\
 20 & -13(28)  \\
 32 & -14(27)  \\
 50 & -18(28)  \\
 64 & -16(28)  \\
 80 & -17(27)  \\
\bottomrule
\bottomrule
\end{tabular}
}
\caption{\label{tab:results_disconnected}  Our results for the disconnected contribution to the squared mass splitting after extrapolation to the physical point for different values of $\Lambda$. }
\end{center}
\end{table}
\begin{figure}
\centering
\includegraphics[width=0.5\linewidth]{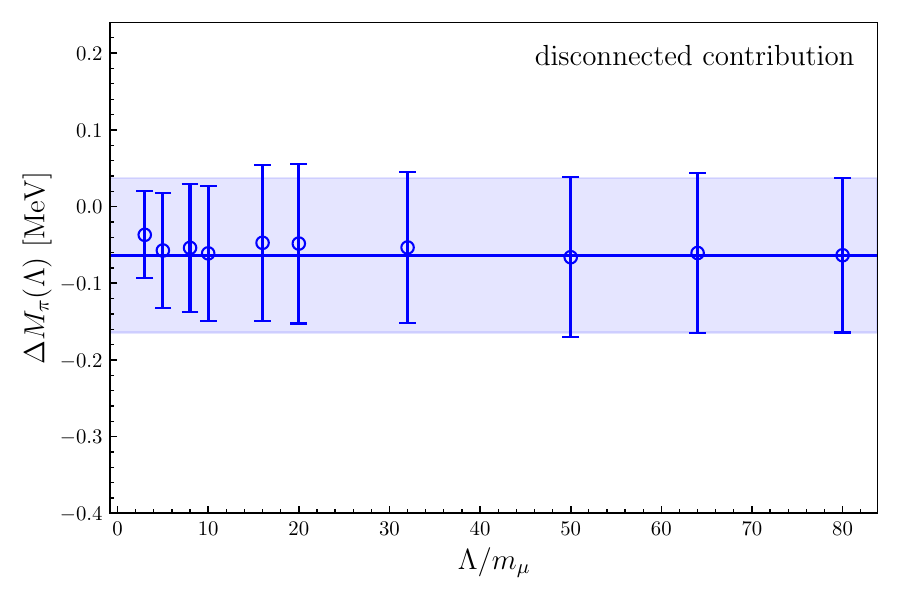}
\caption{\label{fig:disconnected_contribution} Disconnected contribution to the pion mass splitting for different values of $\Lambda$. The horizontal band marks our determination corresponding to $\Lambda=\infty$. }
\end{figure}
To address the dependence on $\Lambda$ we divide the difference of the squared masses by $2M_\pi^\mathrm{phys}$ to obtain $\Delta M_\pi(\Lambda)$. The disconnected contribution to the pion mass splitting thus obtained is shown in Figure~\ref{fig:disconnected_contribution}. We observe that the error on our points is too large to distinguish any dependence on $\Lambda$ and, therefore, we use the value at the highest $\Lambda$ to quote the disconnected contribution in the $\Lambda \to \infty$ limit, obtaining
\begin{flalign}\label{eq:final_disco}
\Delta M_\pi^\mathrm{disc}(\Lambda=\infty)\simeq \Delta M_\pi^\mathrm{disc}(\Lambda=80\,m_\mu)=-0.06(10)\;\text{MeV}.
\end{flalign}
Even though the relative error is above 100\%, our direct computation confirms that the disconnected contribution is $\order{1\%}$ of the connected one and therefore it plays a small role in pion mass splitting.

\section{$\Lambda \to \infty$ limit before the extrapolation to the physical point\label{app:material}}
In the main text of this work, we follow the strategy of first performing the continuum limit, $a \to 0$, and the extrapolation to the physical point, $M_\pi \to M_\pi^\mathrm{phys}$, and only subsequently taking the $\Lambda \to \infty$ limit. The reason for taking the limits in this order originates from the observation that the condition $a\Lambda \ll 1$ must, in general, be satisfied. Such a condition can be satisfied for arbitrarily large values of $\Lambda$ only in the limit $a \to 0$.

We also noted that the pion mass splitting is UV-finite. In this case, the order of the limits can be interchanged without introducing large cutoff effects, as shown in Section~\ref{sec::chiral} of the main text. Relying on this property, we illustrate in this appendix an alternative procedure in which the extrapolation to $\Lambda = \infty$ is performed at fixed lattice spacing, and the chiral-continuum limit is taken only as a final step.

We begin by collecting the results for the pion mass splitting for each ensemble at all values of $\Lambda$ considered in this work. These results, reported in Table~\ref{tab:all_numbers}, are obtained following the procedure described in Section~\ref{sec::long_distance}, in which finite-volume corrections have already been applied.
\begin{table}[t]
\begin{center}
\footnotesize{
\begin{tabular}{c|cccccccc}
\toprule
$\Lambda/m_\mu$ & H102     & H105     & N101     & C101     & S400     & N452     & N451     & D450     \\
\midrule
3    &    0.44(1)  &    0.46(2)   &    0.46(1)  &    0.48(1)  &    0.44(1)  &    0.45(1)  &    0.45(1)  &    0.46(1)  \\
5    &    0.70(1)  &    0.74(3)   &    0.73(2)  &    0.78(2)  &    0.69(1)  &    0.70(1)  &    0.72(1)  &    0.75(2)  \\
8    &    1.00(2)  &    1.08(5)   &    1.08(3)  &    1.17(3)  &    1.00(2)  &    1.02(2)  &    1.05(2)  &    1.12(3)  \\
10   &    1.17(3)  &    1.27(6)   &    1.26(4)  &    1.38(3)  &    1.16(2)  &    1.19(3)  &    1.23(3)  &    1.33(4)  \\
16   &    1.52(4)  &    1.68(8)   &    1.66(5)  &    1.83(4)  &    1.52(3)  &    1.55(4)  &    1.62(4)  &    1.77(5)  \\
20   &    1.68(4)  &    1.86(8)   &    1.83(6)  &    2.04(5)  &    1.68(3)  &    1.71(4)  &    1.80(4)  &    1.98(6)  \\
32   &    1.96(5)  &    2.19(10)  &    2.14(7)  &    2.40(6)  &    1.96(4)  &    1.99(5)  &    2.11(5)  &    2.34(7)  \\
50   &    2.15(6)  &    2.41(11)  &    2.36(8)  &    2.66(7)  &    2.15(4)  &    2.17(5)  &    2.32(5)  &    2.58(8)  \\
64   &    2.23(6)  &    2.50(12)  &    2.44(8)  &    2.76(7)  &    2.22(4)  &    2.25(6)  &    2.40(6)  &    2.68(8)  \\
80   &    2.29(6)  &    2.57(12)  &    2.51(8)  &    2.84(7)  &    2.27(5)  &    2.30(6)  &    2.46(6)  &    2.73(15)  \\
\midrule
$\Lambda/m_\mu$ & D452     & N203     & N200     & D251     & D200     & N302     & J303     \\
\midrule\midrule
3    &    0.55(2)  &    0.44(1)  &    0.45(1)  &    0.45(1)  &    0.48(1)  &    0.45(1)  &    0.45(1)  \\
5    &    0.95(3)  &    0.69(1)  &    0.71(1)  &    0.71(1)  &    0.79(2)  &    0.70(1)  &    0.72(2)  \\
8    &    1.49(5)  &    0.99(2)  &    1.04(2)  &    1.03(2)  &    1.19(3)  &    1.02(2)  &    1.06(3)  \\
10    &    1.79(6)  &    1.16(2)  &    1.22(2)  &    1.20(3)  &    1.41(3)  &    1.19(2)  &    1.25(4)  \\
16    &    2.47(8)  &    1.51(3)  &    1.61(3)  &    1.58(4)  &    1.89(5)  &    1.56(3)  &    1.65(5)  \\
20    &    2.77(9)  &    1.67(4)  &    1.78(4)  &    1.75(4)  &    2.11(5)  &    1.72(3)  &    1.82(6)  \\
32    &    3.32(11)  &    1.95(4)  &    2.09(4)  &    2.04(5)  &    2.50(7)  &    2.00(4)  &    2.14(7)  \\
50    &    3.70(13)  &    2.13(5)  &    2.29(5)  &    2.23(6)  &    2.76(7)  &    2.19(5)  &    2.34(8)  \\
64    &    3.86(13)  &    2.20(5)  &    2.37(5)  &    2.31(6)  &    2.86(8)  &    2.26(5)  &    2.42(8)  \\
80    &    3.98(14)  &    2.25(5)  &    2.42(5)  &    2.36(6)  &    2.93(8)  &    2.31(5)  &    2.47(8)  \\
\bottomrule
\bottomrule
\end{tabular}
}
\caption{\label{tab:all_numbers} 
$\Delta M_\pi(\Lambda)$ in MeV, with the associated statistical uncertainty, for each ensemble and each value of $\Lambda$. These quantities have already been corrected for finite-volume effects, as detailed in Section~\ref{sec::long_distance}. The splitting refers to the connected diagram of Eq.~\eqref{eq:lattice_splitting}. In order to reproduce the statistical correlations between the splittings at different values of $\Lambda$ for a given ensemble, the correlation matrix provided in Eq.~\eqref{eq:correlation} should be used.
}
\end{center}
\end{table}
\begin{figure}
\centering
\includegraphics[width=0.48\linewidth]{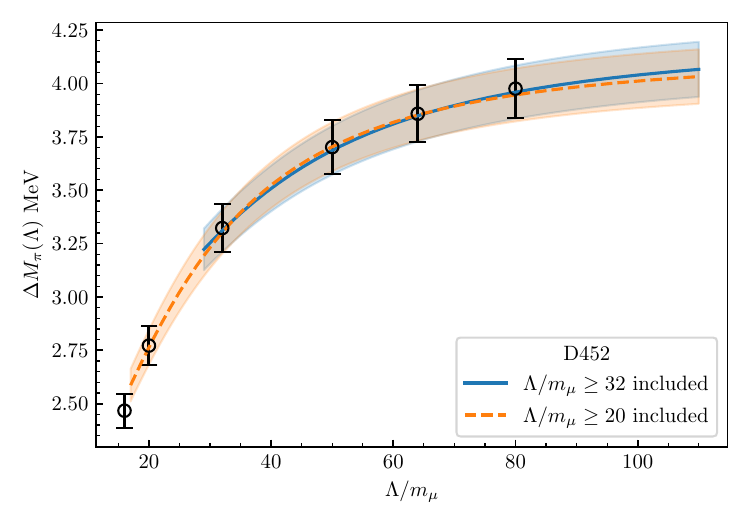}
\includegraphics[width=0.48\linewidth]{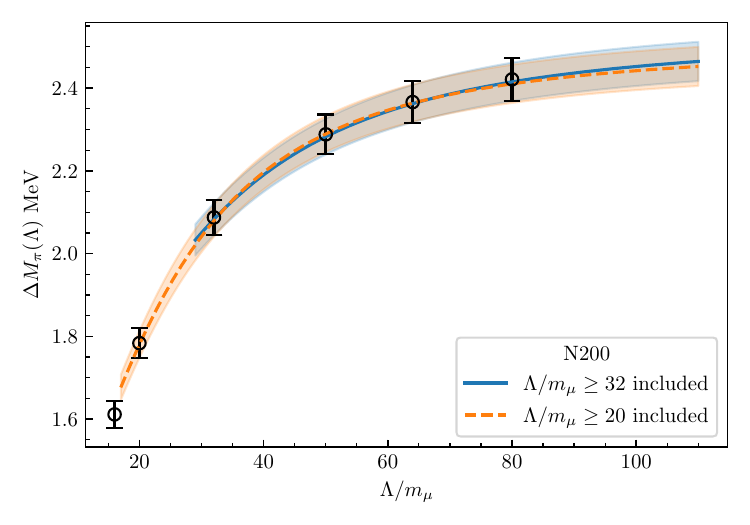}
\caption{\label{fig:lambda_inf_extrapolation}
Extrapolation to $\Lambda = \infty$ obtained by fitting the data points at fixed ensemble D452 (left plot) and N200 (right plot). The solid line shows the result of the fit using only the four points satisfying $\Lambda/m_\mu \ge 32$, while the dashed line corresponds to the fit obtained by including an additional point at $\Lambda = 20\,m_\mu$. The difference between the two fits is taken as an estimate of the systematic uncertainty.
}
\end{figure}

We extrapolate $\Delta M_\pi(\Lambda)$ to $\Lambda = \infty$ separately for each ensemble using the following fit ansatz,
\begin{flalign}\label{eq:ansatz_lambda}
\Delta M_\pi(\Lambda) = c_\infty + \frac{c_1}{\Lambda^2 + c_2},
\end{flalign}
where $c_\infty$, $c_1$, and $c_2$ are parameters determined by fitting directly to the data points. After the fit, the value corresponding to $\Lambda = \infty$ is given by the parameter $c_\infty$. The ansatz in Eq.~\eqref{eq:ansatz_lambda} reflects the $1/\Lambda^2$ behavior induced by the PV-regulated photon propagator and is specifically constructed to approach a constant in the $\Lambda \to \infty$ limit.

As shown in Figure~\ref{fig:lambda_inf_extrapolation} for the ensembles D452 and N200, this ansatz provides a good description of the $\Lambda$ dependence of $\Delta M_\pi(\Lambda)$ at large values of $\Lambda$. In particular, the three fit parameters are determined by fitting the four points corresponding to $\Lambda/m_\mu = 32$, $50$, $64$, and $80$. The fitted value of $c_\infty$, together with its statistical uncertainty, is then quoted as the estimate of $\Delta M_\pi(\Lambda = \infty)$.

To better control the extrapolation, we repeat the fit including an additional point at $\Lambda = 20\,m_\mu$, and take the spread between the two results as an estimate of the systematic uncertainty, which is then combined in quadrature with the statistical one. The results of this procedure are reported for each ensemble in Table~\ref{tab:numbers2}.
\begin{table}[t]
\begin{center}
\footnotesize{
\begin{tabular}{c|c}
\toprule
ID & $\Delta M_\pi$ [MeV] \\
\midrule
\midrule
H102  &   2.40(6)(3)[7]   \\
H105  &   2.70(12)(4)[12] \\
N101  &   2.64(8)(4)[9]   \\
C101  &   3.00(7)(5)[9]   \\
\midrule
S400  &   2.38(4)(3)[5]   \\
N452  &   2.40(6)(3)[6]   \\
N451  &   2.58(6)(3)[6]   \\
D450  &   2.85(18)(3)[18] \\
D452  &   4.20(14)(7)[15] \\
\midrule
N203  &   2.35(5)(2)[5]   \\
N200  &   2.53(5)(2)[5]   \\
D251  &   2.46(6)(2)[6]   \\
D200  &   3.06(8)(3)[8]   \\
\midrule
J303  &   2.57(8)(2)[8]   \\
N302  &   2.40(5)(2)[5]   \\
\bottomrule
\bottomrule
\end{tabular}
}
\caption{\label{tab:numbers2} Pion mass splitting for each ensemble after the $\Lambda\to\infty$ extrapolation.  The first error is statistical, the second systematic and the third is the total error obtained as the combination in quadrature of the first two.}
\end{center}
\end{table}

We note that the elastic contribution, evaluated using the Cottingham formula presented in Section~\ref{sec::pheno}, and employing a parametrization of the electromagnetic form factor based on the VMD masses reported in Table~\ref{tab:ensembles}, can in principle also be determined at the level of each individual ensemble. Subtracting this elastic contribution from the total mass splitting would then allow one to isolate the inelastic part and perform the $\Lambda \to \infty$ extrapolation following an alternative strategy, analogous to that used in Section~\ref{sec::pheno}.

We do not pursue this approach here, as it is less phenomenologically relevant in the present context and the method described above is sufficient to control the $\Lambda \to \infty$ limit. We simply note that, for mass splittings corresponding to $M_\pi \ge 220~\text{MeV}$, the inelastic contribution is found to be negative.

The final step in obtaining the physical pion mass splitting is the extrapolation to the continuum and to the physical point. To this end, we follow the same procedure as in the main text and convert the pion mass splittings reported in Table~\ref{tab:numbers2} into the squared difference, $M_{\pi^+}^2 - M_{\pi^0}^2$, by multiplying each value by $2 M_\pi$.

We then perform a simultaneous fit of the dependence on $a$ and $M_\pi$ using the fit ansatz of Eq.~\eqref{eq:fit_ansatz}. To carry out the extrapolation and assign a systematic uncertainty, we repeat the analysis described in Section~\ref{sec::chiral}, considering three different fit variants: setting $c_a = c_{a\pi} = 0$, setting $c_{a\pi} = 0$, and imposing no constraints on these parameters.

The fits are further repeated by excluding selected data points, one at a time, resulting in a total of 19 fits, which are then combined according to the AIC procedure described in Section~\ref{sec::chiral}. The systematic uncertainty is determined from the weighted average defined in Eq.~\eqref{eq:systematic_error}.
\begin{figure}[t]
\centering
\includegraphics[width=0.93\linewidth]{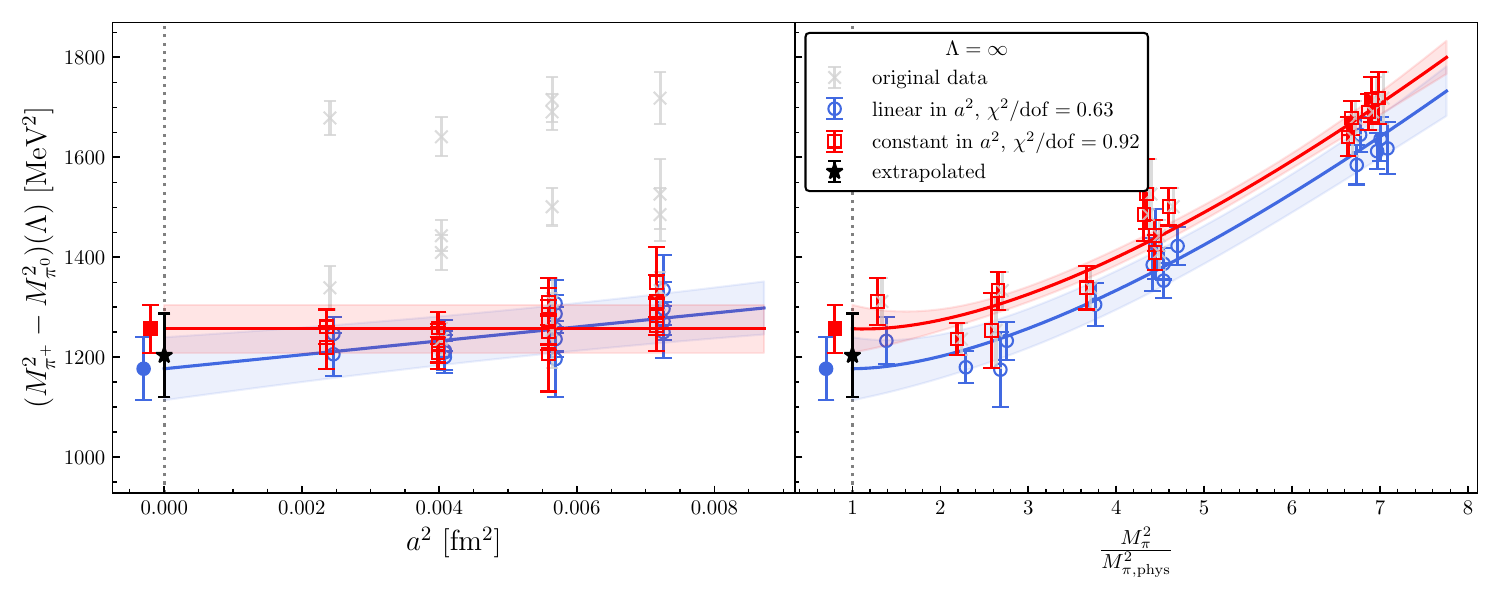}
\caption{\label{fig:chiral_cont_alternative}  Chiral and continuum extrapolation of the squared pion mass splitting after extrapolating each point to $\Lambda=\infty$.  See Figure~\ref{fig:chiral_cont} for additional details. The black point, marked by a star, is the final result of the extrapolation already including a systematic error.}
\end{figure}
The results of the fits including all data points, assuming constant and linear dependence on $a^2$, respectively, are shown in Figure~\ref{fig:chiral_cont_alternative}, which is fully equivalent to Figure~\ref{fig:chiral_cont} (see its caption for reference). As can be seen, the extrapolation is well controlled and, compared to the analysis in Section~\ref{sec::chiral}, which is performed at finite $\Lambda$, there is a slight preference for the fit including the $a^2$ term. This observation is consistent with the trend shown in Figure~\ref{fig:weights}, where an increasing sensitivity to lattice cutoff effects is observed as $\Lambda$ increases.

The final result obtained within this alternative analysis strategy reads
\begin{flalign}
[M_{\pi^+}^2 - M_{\pi^0}^2] = 1203(56)(62)[84]\;\text{MeV}^2.
\end{flalign}
Dividing this by $2 M_\pi^\mathrm{phys}$, and adding the disconnected contribution from Eq.~\eqref{eq:final_disco}, yields
\begin{flalign}
\Delta M_\pi = 4.40(24)(23)[33]\;\text{MeV},
\end{flalign}
which is compatible with the determination of Eq.~\ref{eq:final_conn}, obtained by performing the $\Lambda \to \infty$ extrapolation after the continuum and physical-point limits, at the level of less than half a standard deviation.

We find that the total uncertainty from this alternative procedure is about 50\% larger than that obtained in the main text. The main reasons can be traced back to the fact that: i) an additional systematic uncertainty is associated with the $\Lambda \to \infty$ extrapolation at fixed ensemble, and ii) the dominant fit in the chiral-continuum extrapolation is the one including the $a^2$ term, which carries a larger statistical uncertainty compared to the fit in which this term is omitted.

This alternative procedure provides a valuable consistency check of the analysis presented in the main text, which we use to quote our final result for the pion mass splitting.
\newpage
\bibliographystyle{JHEP}
\bibliography{references}
\end{document}